\documentclass[10pt,twocolumn]{article}

\usepackage[
    paper=a4paper,
    textheight=235mm, 
    textwidth=177mm,  
    top=20mm,         
    bottom=20mm,      
    bindingoffset=0mm
]{geometry}

\usepackage{amsmath,amsfonts,amsthm} 
\usepackage{mathrsfs}
\usepackage{xfrac}
\usepackage{physics}
\usepackage{siunitx}
\usepackage[switch]{lineno}
\providecommand{\Newlabel}[2]{}
\usepackage[T1]{fontenc}
\usepackage{newtxtext} 
\usepackage{newtxmath} 
\usepackage{microtype}

\usepackage{graphicx}
\usepackage{multirow}
\usepackage{booktabs}
\usepackage{threeparttable}

\usepackage{algorithm}
\usepackage{algorithmicx}
\usepackage{algpseudocode}
\usepackage{listings}

\usepackage[title]{appendix}
\usepackage{xcolor}
\usepackage{textcomp}
\usepackage{manyfoot}
\usepackage[normalem]{ulem}
\usepackage{enumitem}

\usepackage[numbers,sort&compress]{natbib}
\usepackage[hidelinks]{hyperref}
\usepackage[nameinlink,capitalise,noabbrev]{cleveref}

\usepackage{titlesec}

\titleformat{\section}
  {\normalfont\rmfamily\bfseries\large}  
  {}
  {0pt}
  {}

\titleformat{\subsection}
  {\normalfont\rmfamily\bfseries\normalsize}
  {}
  {0pt}
  {}

\newcommand{\ecat}{\ket{\mathcal{C}_\alpha^+}}
\newcommand{\ecata}[1]{\ket{\mathcal{C}_{#1}^+}}
\newcommand{\ocat}{\ket{\mathcal{C}_\alpha^-}}
\newcommand{\ocata}[1]{\ket{\mathcal{C}_{#1}^-}}
\newcommand{\execat}{|\tilde{\mathcal{C}}_\alpha^-\rangle}
\newcommand{\exocat}{|\tilde{\mathcal{C}}_\alpha^+\rangle}

\usepackage{titlesec}

\usepackage{caption}
\title{\fontsize{13pt}{16pt}\fontseries{b}\selectfont Motional Kerr-Cat States of an Atom in an Optical Tweezer}
\author{
    \normalsize Steven Pampel$^{\dagger1,2}$, Gur Lubin$^{\dagger1,2}$, Dawson P. Hewatt$^{1,2}$, Conall McCabe$^{1,2,3}$, Jaeyong Hwang$^{1,2,3}$, \\
    \normalsize Sean R. Muleady$^{4,5}$, Tianrui Xu$^{6}$, Ana Maria Rey$^{1,2,3}$, and Cindy A. Regal$^{*1,2}$
}
\date{}

\begin{document}
\rmfamily

\twocolumn[
  \begin{@twocolumnfalse}
    \maketitle
    
    \begin{center}
        \vspace{-2em} 
        \normalsize 
        $^{1}$JILA, NIST and University of Colorado, Boulder, Colorado 80309, USA\\
        $^{2}$Department of Physics, University of Colorado, Boulder, Colorado 80309, USA\\
        $^{3}$Center for Theory of Quantum Matter, University of Colorado, Boulder, Colorado 80309, USA\\
        $^{4}$Joint Center for Quantum Information and Computer Science, NIST and University of \\ Maryland, College Park, Maryland 20742, USA\\
        $^{5}$Joint Quantum Institute, NIST and University of Maryland, College Park, Maryland 20742, USA\\
        $^{6}$Institut Quantique, Université de Sherbrooke, Québec J1K 2R1, Canada

        \noindent${}^{\dagger}$ these authors contributed equally to this work

    \end{center}
    
    \vspace{2.5em}

    \begin{center}
    \begin{minipage}{0.85\textwidth} 
    \normalsize 
    \noindent
Schrödinger cat states --- quantum superpositions of classically or macroscopically distinct states --- constitute a powerful resource for quantum computing~\cite{aliferis2008fault,grimm2020stabilization,putterman2025preserving,putterman2025hardware,qing2026quantum}, enhanced metrology~\cite{facon2016sensitive,cao2024multi,yang2025minute,Kasevich1991,Greve2022Entanglement,Cassens2025Entanglement}, and probing coherence on large scales~\cite{brune1996,fein2019quantum,pedalino2026probing}. Encoding such states in the phase space of an oscillator requires a nonlinearity~\cite{yurke1986generating,milburn1986quantum}, typically inherited from an auxiliary degree of freedom such as atomic spin~\cite{monroe1996schrodinger,Lo2015,johnson2017ultrafast} or a Josephson junction~\cite{friedman2000,vlastakis2013,leghtas2015confining,puri2019cat,bild2023schrodinger}. Neutral atoms in optical tweezer arrays — a leading platform for quantum science and computing~\cite{kaufman2021quantum,henriet2020quantum} — provide an intrinsic nonlinearity via the motion of a single atom in a tightly focused trap. However, this self-Kerr mechanism has not previously been exploited for cat-state generation, and remains largely unexplored as a resource for motional-state control.  Here we realize Schrödinger cat states in the quantized motion of a single neutral atom trapped in an optical tweezer. By modulating the trap depth and position, we demonstrate parity control of both Kerr-cat and Fock states alongside tunable nonlinearity, establishing a spin- and species-independent framework for controlling motion. We further show that the cat-state encoding is intrinsically robust against trap-frequency fluctuations that otherwise limit the fidelity of direct Fock-state transitions. These results establish Kerr-based control of neutral-atom motion as a new paradigm for cat-state and bosonic-state engineering in optical tweezers, providing a route toward quantum-error-correcting codes such as grid states~\cite{fluhmann2019encoding,campagne2020quantum}, and toward quantum-enhanced sensing with arrays of non-Gaussian states~\cite{mccormick2019quantum,Wolf2019Motional}.
    \end{minipage}
    \end{center}

    \vspace{2em} 
  \end{@twocolumnfalse}
]

\maketitle

\begin{figure*}[t]
\centering
\includegraphics[width=\textwidth]{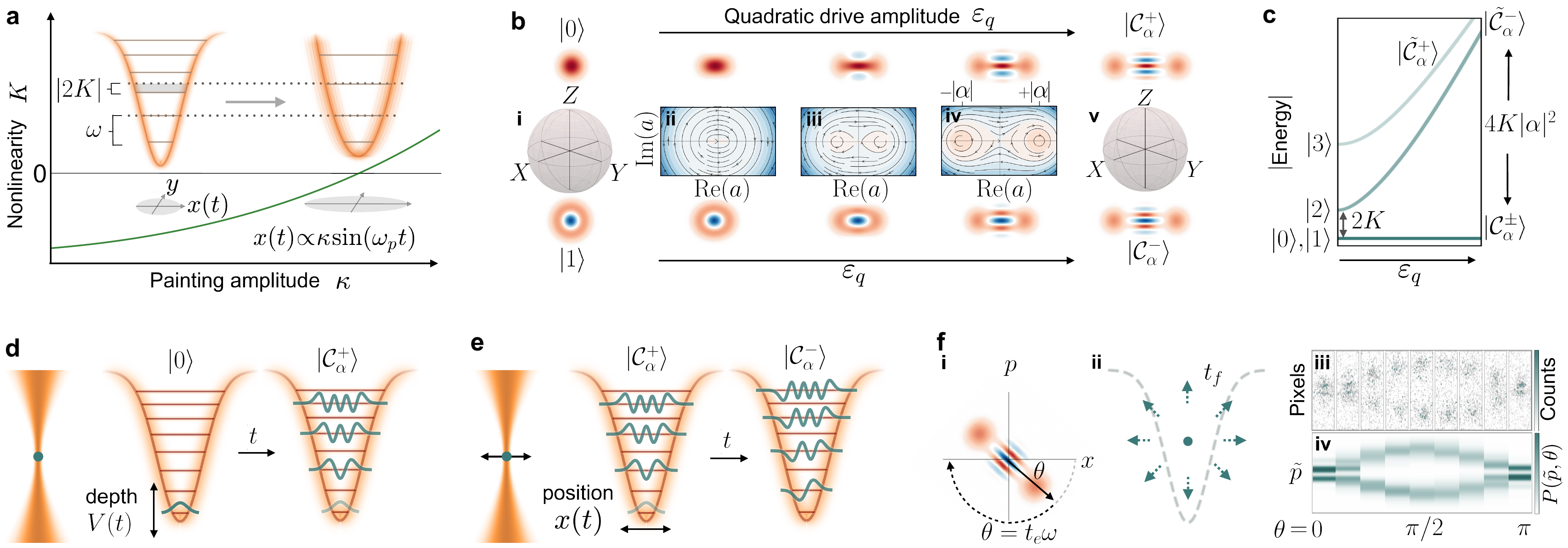 }\caption{\textbf{Motional Kerr-cat state preparation and tomography in optical tweezers.}
\textbf{a,} Kerr nonlinearity $K$ versus painting amplitude $\kappa$ in a tweezer with trap frequency $\omega$. All parameters are defined along the $x$-dimension of the tweezer, where $(x,y)$ and $z$ denote the radial and axial dimensions, respectively. Modulation of the trap position at a frequency $\omega_p \gg\omega$ (painting) and amplitude $\kappa$ enables dynamic tuning of the anharmonicity $2K$ between the intrinsic Gaussian value and the harmonic limit. Shaded ellipses indicate the trap waist in $x$ and $y$. \textbf{b,} Adiabatic Kerr-cat preparation. An initially prepared Fock state $|0\rangle$ or $|1\rangle$ (\textbf{i}) is transformed into an even- or odd-parity Kerr-cat state $|\mathcal{C}_\alpha^{+}\rangle$ and $|\mathcal{C}_\alpha^{-} \rangle$ (\textbf{v}) by adiabatically increasing the quadratic drive amplitude $\varepsilon_q$.  Classical quasienergy landscapes in the rotating frame (\textbf{ii-iv}) are shown at different values of $\varepsilon_q$, where $[\mathrm{Re}(a),\mathrm{Im}(a)]$ are the rotating-frame phase-space coordinates. As $\varepsilon_q$ increases, the initial landscape is first squeezed (\textbf{ii}) and then bifurcates into a double-well with minima at $\pm|\alpha|$ (\textbf{iii–iv}). Simulated Wigner functions for the corresponding even- and odd-parity states are shown above and below the quasienergy landscapes. \textbf{c,} Eigenenergies in a frame rotating at $\omega$ as a function of $\varepsilon_q$. Slowly increasing $\varepsilon_q$ enhances the energy gap between ground- and excited-state manifolds from $2K$ to $4K|\alpha|^2$. ($|\tilde{\mathcal{C}}_\alpha^\pm\rangle$ are the first-excited cat states.) \textbf{d,} Quadratic drive for preparing even-parity cat states. Modulation of the trap depth $V(t)$ at frequency $\sim2\omega$ couples states of the same parity, allowing adiabatic mapping between the ground Fock state $\ket{0}$ and $|\mathcal{C}_{\alpha}^{+}\rangle$. \textbf{e,} Linear drive for preparing odd-parity cat states. Modulation of trap position $x(t)$ at a frequency $\sim\omega$ couples states of opposite parity, allowing rotations between $|\mathcal{C}_{\alpha}^{+}\rangle$ and $|\mathcal{C}_{\alpha}^{-}\rangle$. \textbf{f,} Time-of-flight motional-state tomography. The atom evolves in phase-space under the Kerr Hamiltonian ($\varepsilon_q\!=\varepsilon_l\!=\!0$) for a time $t_e=\theta/\omega$ (\textbf{i}), where $\theta$ is the phase-space angle and $(x,p)$ are position and momentum coordinates. Ballistic expansion for a time $t_f$ (\textbf{ii}) maps the in-trap momentum distribution at a given $\theta$ onto a spatial distribution (\textbf{iii}), from which the maximum-likelihood estimate of the probability density $P(\tilde p,\theta)$ is obtained (\textbf{iv}). (Note that panel \textbf{i} displays a simulated Wigner function, whereas \textbf{iii} and \textbf{iv} show experimental data; additional details are provided in \autoref{fig:cats}d.)}
\label{fig:intro_figure}
\end{figure*}

Schrödinger cat states have emerged as an indispensable resource for applications ranging from quantum error correction~\cite{aliferis2008fault,grimm2020stabilization,putterman2025preserving,putterman2025hardware,qing2026quantum} and metrology~\cite{parazzoli2012observation,steffen2012digital,facon2016sensitive,omran2019generation,cao2024multi,yang2025minute,zheng2026quantum} to studies of mesoscopic superpositions~\cite{brune1996,fein2019quantum,pedalino2026probing}. Unlike cat states encoded in free-space trajectories or discrete internal degrees of freedom, Kerr-cat states reside in the infinite-dimensional Hilbert space of an oscillator with Kerr nonlinearity (\autoref{fig:intro_figure}a). Here, a parametric drive stabilizes superpositions of displaced ground states $\ket{\pm\alpha}$ whose even- and odd-parity combinations $|\mathcal{C}_{\alpha}^{\pm}\rangle \propto \ket{\alpha}\pm\ket{-\alpha}$ provide a protected logical subspace for qubit operations~\cite{cochrane1999macroscopically} (\autoref{fig:intro_figure}b,c) and a sub-Planck interference structure for quantum-enhanced displacement sensing~\cite{zurek2001sub,toscano2006sub}. 

In optical tweezer arrays, where tightly-focused light confines individual neutral atoms within independently controlled traps, the quantized motion of an atom offers a natural, yet untapped, degree of freedom for realizing Kerr-cat states in a  reconfigurable environment. This opportunity arises from the geometric anharmonicity --- or self-Kerr nonlinearity --- of the Gaussian trapping potential, where adjacent energy-level spacings deviate from those of a harmonic oscillator by several percent~\cite{cortinas2024towards,grochowski2025quantum,lunt2024realization,serwane2011deterministic}. Consequently,  motional eigenstates can be selectively addressed  without spin-coupled Raman-sideband transitions, extending coherent control  to shallow traps where recoil heating is negligible and programmable inter-site tunneling can be exploited~\cite{kaufman2015entangling,gonzalez2023fermionic,brown2023time}. While nascent compared to  optical lattices~\cite{gross2017quantum,hartke2022quantum}, such capabilities are increasingly vital in dynamic tweezer arrays as motional Hamiltonian engineering  emerges as a powerful frontier. Applications include continuous-variable quantum information processing~\cite{shaw2025erasure,bohnmann2025bosonic}, suppression of motion-induced errors in Rydberg gates~\cite{levine2018high,lienhard2025generation}, itinerant quantum simulation~\cite{kaufman2015entangling,murmann2015two,spar2022realization}, and programmable logical processors with native fermionic encodings~\cite{gonzalez2023fermionic}.

In this work, we demonstrate the generation and parity control of Kerr-cat and Fock states via modulation of an anharmonic optical tweezer (\autoref{fig:intro_figure}d,e). We highlight key advantages of cat states over Fock states: faster parity rotations and enhanced robustness to trap-frequency fluctuations. Using time-of-flight (TOF) quantum-state tomography~\cite{brown2023time} (\autoref{fig:intro_figure}f), we measure cat states with fidelities comparable to those achieved in state-of-the-art superconducting Kerr-cat platforms~\cite{grimm2020stabilization,putterman2025hardware,qing2026quantum}. More broadly, these results establish Kerr nonlinearity of optical tweezers as a general resource for preparing non-Gaussian states, applicable across atomic and molecular species, and potentially levitated nanoparticles~\cite{delic2020cooling}.

\subsubsection{Tuning Anharmonicity in Tweezers}\label{subsec:kerr}

Precise control over an oscillator's nonlinearity enables tailored engineering of its effective Hilbert space. While strong nonlinearity introduces non-degenerate energy spacings that enable selective manipulation of individual Fock states, a weaker nonlinearity permits coherent population transfer across multiple levels, facilitating the generation of cat states and squeezed states  ~\cite{lienhard2025generation}.  To bridge these two regimes, we exploit the tunable geometry of an optical tweezer potential, retaining selective Fock-state control in the native Gaussian trap while reducing anharmonicity to access higher-lying motional levels for cat-state preparation.

Our starting point is a single $^{87}$Rb atom prepared near its three-dimensional motional ground state (see Methods\ref{met:timing}) in an approximately Gaussian optical tweezer. Unless otherwise stated, we focus exclusively on the atom's motion in the $x$-dimension of a tweezer with radial (axial) coordinates $x,y$ ($z$). While this motion is nearly harmonic when the extent of the atomic wave packet remains small compared to the trap waist, quartic corrections to the Gaussian potential shift the $n$-th motional eigenfrequency as
$\omega_n \!\approx \!n[\omega  + K (n - 1)]$. Here, $\omega \equiv \omega_1\!-\!\omega_0$ is the trap frequency and $K$ is the Kerr nonlinearity, which can be expressed in terms of the absolute anharmonicity, 
\begin{equation}
    2K = (\omega_2 - \omega_1) - (\omega_1 - \omega_0),    
\end{equation} 
or as a dimensionless relative anharmonicity $\eta \equiv {2K}/{\omega}$.
For a Gaussian potential, $2K \!\approx \!-{3\hbar}/{(2mW^2)}$ and $\eta \!\approx\!-3(x_0/W)^2$, where $x_0=\sqrt{\hbar/2 \omega m}$ is the oscillator length, $W$ is the trap waist, and $m$ is the atom's mass. By sinusoidally modulating the tweezer position such that $x \rightarrow x -\kappa W\sin(\omega_pt)$ at a frequency $\omega_p\gg\omega$ and dimensionless amplitude $\kappa$, the atom experiences a time-averaged, or  ``painted'',  potential. For $\kappa\lesssim 0.5$ the anharmonicity is well-approximated by (see Methods\ref{met:painted_tweezer_theory})

\begin{equation}
     \eta\approx -3\left(\frac{x_0}{W}\right)^2(1-2\kappa^2).  
\end{equation}
Thus, $\eta$ can be tuned continuously from the native Gaussian value  of $\eta \!\approx\! -5\%$ (for $\omega/2\pi\!\approx\!\SI{8}{kHz}$ and $W\approx\SI{0.7}{\micro m}$)
to that of a nearly harmonic trap with $\eta \!\approx\!0$, and even into a regime of a spatial double-well (\autoref{fig:intro_figure}a).

\begin{figure}[t!]
\centering
\includegraphics[width=\columnwidth]{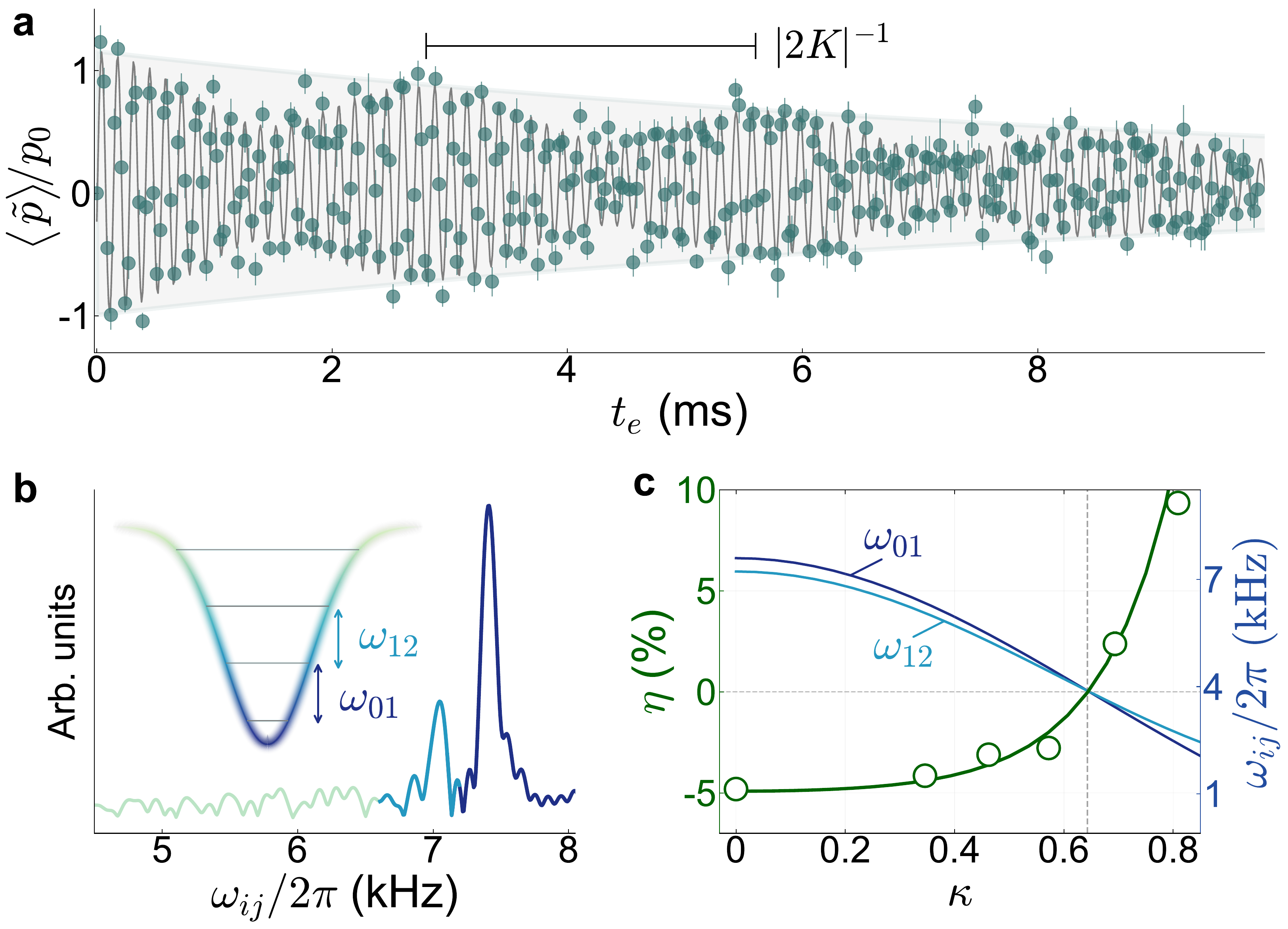}
\caption{
\textbf{Tuning and measuring Kerr nonlinearity in an optical tweezer. a,} Momentum expectation value $\langle \tilde p \rangle/p_0$, where $p_0=\sqrt{m\hbar \omega/2}$ is the zero-point momentum, as a function of evolution time $t_e$ for an atom in a displaced motional ground-state of an oscillator with Kerr nonlinearity $K$. Grey lines and shading represent a fit to the sum of two sinusoids with an exponentially decaying envelope (see Methods\ref{met:coherent_state}). \textbf{b,} Fourier spectrum of the time-domain data shown in panel (a), revealing motional transition frequencies $\omega_{12} \equiv \omega_2 \!-\! \omega_1$ and $\omega_{01} \!\equiv \!\omega_1 \!-\! \omega_0 \!=\!\omega$. \textbf{c,} Relative anharmonicity $\eta\!=\!{2K/\omega}$ (left axis) and motional eigenfrequencies (right axis) as a function of the painting amplitude $\kappa$, showing tunability between a Gaussian ($\kappa \!=\!0$) and an approximately harmonic potential ($\kappa \!\approx\!0.6$). Circles are experimental data and lines are simulated (see Methods\ref{met:coherent_state}). For $\kappa \!\gtrsim \!0.6$, anharmonicity becomes positive as the potential approaches a spatial double-well (not used in this work). Error bars in panel (a) denote $1\sigma$ uncertainties of the fits; errors in panel (c) are smaller than the marker size.}
\label{fig:kerr_figure}
\end{figure}

We characterize $\eta$ as a function of $\kappa$ by observing the phase-space dynamics of a displaced ground state. Using TOF imaging, we measure the expectation value of the momentum quadrature $ \tilde p(\theta \!=\! t_e \omega) \!=\! p \cos\theta \!+\! \left(p_0/{x_0}\right) x \sin\theta$, where $t_e$ is the phase-space evolution time and $p_0=\sqrt{m\hbar \omega/2}$ is the zero-point momentum. Fitting the time-domain signal (\autoref{fig:kerr_figure}a) to a sum of two sinusoids with an exponentially decaying envelope reveals the motional transition frequencies $\omega_{12} \equiv \omega_2 \!-\! \omega_1$ and $\omega_{01}=\omega$ (\autoref{fig:kerr_figure}b). Repeating this measurement for different values of $\kappa$ reveals its relationship to $\eta$ (\autoref{fig:kerr_figure}c), in good agreement with numerical simulations (see Methods\ref{met:coherent_state}). 

While painting enables precise tuning of $K$, the technique also provides flexible control over the two-dimensional radial mode structure, allowing the degeneracy between the radial tweezer axes ($x,y$) to be lifted or maximized. In the former case, painting along a single radial axis generates an elliptical profile that suppresses cross-mode coupling. This ensures that the motional dynamics remain effectively one-dimensional, thereby improving state fidelity in the dimension of interest while leaving the anharmonicity unchanged. Therefore, we paint along the $y$-dimension when preparing motional states in potentials that are unpainted along $x$.

\begin{figure*}[t]
\centering
\includegraphics[width=\textwidth]{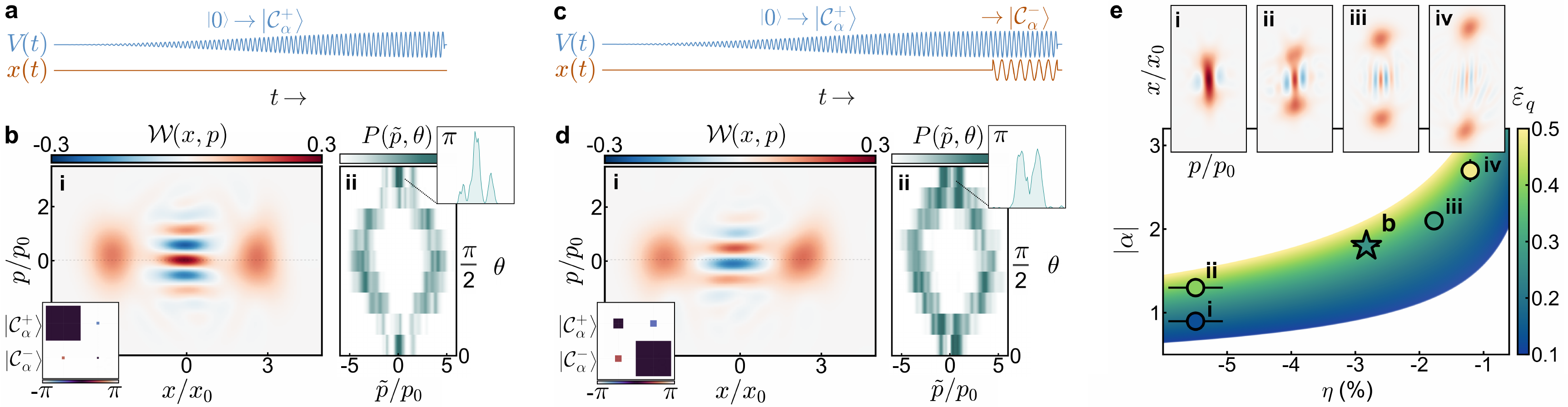}
\caption{\textbf{$|$ Motional Kerr-cat states of a single atom.}
\textbf{a,c,} Pulse sequences for preparing even- and odd-parity cat states $|\mathcal{C}_\alpha^{\pm}\rangle$. An adiabatic, frequency-chirped modulation of the trap depth $V(t)$ transfers an atom from its motional ground state $|0\rangle$ to $|\mathcal{C}_\alpha^+\rangle$ (a). Subsequent modulation of the trap position $x(t)$ rotates $|\mathcal{C}_\alpha^+\rangle$ to  $|\mathcal{C}_\alpha^-\rangle$ (c).
\textbf{b,d,} Tomographic reconstruction of $|\mathcal{C}_{1.8}^{+}\rangle$ (b) and $|\mathcal{C}_{1.8}^{-}\rangle$ (d) indicate state fidelities $\num{94.4(2.9:2.7)}\%$ and $\num{74.8(1.9:1.9)}\%$, respectively (uncertainties are 68.2\% bootstrap confidence intervals). Corresponding Wigner functions (\textbf{i}) and Hinton plots (insets in \textbf{i}) are reconstructed from TOF quadrature distribution measurements $\tilde{p}/p_0$ (\textbf{ii}). The interference fringes in the probability density $P(\tilde p,\theta)$ at $\theta=\pi$ (insets in \textbf{ii}) reveal the state parity. 
\textbf{e,}~Cat size $\abs{\alpha}$ versus relative anharmonicity $\eta\equiv2K/\omega$, where $\omega/2\pi\!\approx \!\SI{8}{kHz}$. Circles indicate experimentally realized even-parity cat states, with their corresponding reconstructed Wigner functions displayed above. The star denotes the state in (b), while the color gradient shows the theoretical prediction (\autoref{eq:alpha}) over the approximate  
range of effective drives $\tilde \varepsilon_q$ explored in this work. Error bars represent $1\sigma$ fit uncertainties, and are visible only where they exceed marker size.}
\label{fig:cats}
\end{figure*}

\subsubsection{Kerr-based control of atomic motion}
\label{sec:encoding}

The motion of an atom in a tweezer can be controlled using a combination of linear and quadratic drives, analogous to the single- and two-photon drives used in transmon-coupled resonators~\cite{grimm2020stabilization,yang2024mechanical}. Unlike motional-state control via tunnel coupling~\cite{brown2023time,wirth2011evidence} or globally driven lattices~\cite{eckardt2017colloquium,sandholzer2022floquet},
 our resonant drives operate on an isolated atom within a single well~\cite{serwane2011deterministic,lunt2024realization}, enabling local
preparation of single-atom motional superpositions. We implement linear and quadratic drives by sinusoidally modulating the tweezer position $x$ and depth $V$, respectively:
\begin{align}
    \label{eq:linear_drive}
    x(t) &= \varepsilon_l x_0 \cos(\omega_l t), \\[6pt] 
    \label{eq:quadratic_drive}
    V(t) &= V_0\left(1 + \varepsilon_q \cos(\omega_q t)\right),
\end{align}
where $V_0$ is the initial trap depth, $\varepsilon_{l,q}$ are the dimensionless drive amplitudes, and $\omega_{l,q}$ are the drive frequencies. When $\omega_l \!\approx \!\omega$ ($\omega_q\approx2\omega$), the linear (quadratic) drive couples motional states of opposite (equal) parity, enabling the transitions $|n\rangle \leftrightarrow |n\pm1\rangle$ ($|n\rangle \leftrightarrow |n\pm2\rangle$) [\autoref{fig:intro_figure}e(d)]. Here, $\ket{n}$ is the $n$-th Fock state of the harmonic oscillator, where $\hat{a}^\dagger \hat{a}\ket{n}= n \ket{n}$ with creation/annihilation operators $\hat{a}^\dagger/\hat{a}$. Substituting $x(t)$ and $V(t)$ into the tweezer potential and transforming to a frame rotating at $\omega_l = \omega_q/2$ yields the effective Hamiltonian (see Methods):
\begin{equation}
\label{eq:KCHamiltonian}
\begin{split}
\hat H /\hbar\approx{} & -\delta \hat{a}^{\dagger}\hat{a} + K \hat{a}^{\dagger2}\hat{a}^2 \\
& + \frac{\omega\varepsilon_q}{8}\left(\hat{a}^{\dagger2}+\hat{a}^2\right) +
\frac{\omega\varepsilon_l}{4}\left(\hat{a}^{\dagger}+\hat{a}\right),
    \end{split}
    \end{equation}
where $\delta \equiv \omega_q/2-\omega$ is the drive detuning.

The Hamiltonian preserves parity in the absence of a linear drive ($\varepsilon_l\!=\!0$), enabling continuous evolution between Fock and cat states when $\varepsilon_q$ is varied adiabatically relative to $|2K|^{-1}$. In this case, the ground state becomes two-fold degenerate and can be written as $ \ket{\mathcal{C}_\alpha^\pm}\propto [\mathcal{ \hat D}(\alpha)\pm \mathcal{\hat D}(-\alpha)]\ket{0}$, where $\hat{\mathcal{D}}(\pm\alpha)=e^{\pm(\alpha \hat{a}^\dagger - \alpha^* \hat{a})}$ is the displacement operator and $|\pm \alpha\rangle\!\equiv\!\hat{\mathcal{D}}(\pm\alpha)|0\rangle$. For large $|\alpha|$, the first-excited parity doublet evolves into
$|\mathcal{\tilde C}_\alpha^\pm\rangle\propto [\mathcal{\hat D}(\alpha)\pm \mathcal{\hat D}(-\alpha)]\ket{1}$, with higher manifolds following analogously~\cite{puri2019cat}. Semiclassically, this evolution corresponds to a bifurcation of the rotating-frame phase-space potential from a single minimum to a double well (\autoref{fig:intro_figure}b) with stable minima located at 
\begin{equation}\label{eq:alpha}
    \pm|\alpha| \approx \pm\sqrt{\frac{\omega\varepsilon_q/8-\delta/2}{\abs{K}}}.
\end{equation}
During this bifurcation, the energy gap isolating the $\ket{\mathcal{C}_\alpha^\pm}$ ground-state manifold  increases from $2K\!-\!2\delta $ to 
$4K |\alpha^2|$ (\autoref{fig:intro_figure}c). Because the gap reaches its minimum value of $2K$ at the onset of the quadratic ramp ($\delta\!=\!0, \ \varepsilon_q\!=\!0$), leakage to excited states can be suppressed by dynamically chirping the quadratic-drive frequency, thereby maintaining a larger instantaneous gap throughout the evolution (see Methods\ref{met:chirping}).

In the presence of a linear drive ($\varepsilon_l \neq 0$), the Hamiltonian couples states of opposite parity, enabling coherent control between the even and odd components of the ground cat-state parity doublet. We define a Bloch sphere with $\ket{\mathcal{C}^\pm_{\alpha}}$ at the poles, the coherent states $\ket{\pm \alpha}$ along its $X$-axis, and $\ket{\mathcal{C}^{\mp i}_{\alpha}}\propto \ket{\alpha} \mp i\ket{-\alpha}
$ along its $Y$-axis (\autoref{fig:intro_figure}b). In this basis, where a linear drive implements rotation in $X$, both parity-rotation Rabi rates and leakage suppression scale favorably for $\abs{\alpha}$. Specifically, the Rabi rate scales as $\Omega= \omega\varepsilon_l |\alpha| / 2$, and the energy-gap increase from $2K$ to $4K|\alpha|^2$ reduces the minimum leakage-free parity-inversion time by a factor of $1/\left(2|\alpha|^2\right)$. Additionally, the linear-drive matrix element that couples $|\mathcal{ C}_{\alpha}^{\pm}\rangle$ and $|\mathcal{\tilde C}_{\alpha}^{\mp}\rangle$ is a factor of $2|\alpha|$ smaller than the matrix element that couples $|\mathcal{ C}_{\alpha}^{+}\rangle$ and $|\mathcal{ C}_{\alpha}^{-}\rangle$, further suppressing leakage (see Methods\ref{met:matrix_Hamiltonians}). Note that rotations in $Z$—which are not explored in this work— can be implemented by detuning the quadratic drive~\cite{qing2026quantum}, but are generally much slower than rotations in $X$ due to a gate time that scales exponentially with $|\alpha|$.

\subsubsection{Realizing Kerr-cat states in tweezers }
\label{subsec:realizing_cat}

Using a painted potential with $\omega/2\pi \approx\SI{8}{kHz}$ and $\eta\!\approx\! -3\%$, we prepare an even-parity cat state by adiabatically turning on a frequency-chirped quadratic drive (see Methods\ref{met:mod_methods} and \autoref{fig:cats}a). TOF quadrature measurements acquired over the full range of phase-space rotation angles $\theta$ show the expected phase-space separation and interference fringes at $\theta\!=\!0,\pi$. The density matrix and corresponding Wigner function (\autoref{fig:cats}b), reconstructed using a maximum likelihood estimation algorithm, indicate an even-parity cat state $\ecata{1.8}$ with fidelity $\mathcal{F}^{+}_{1.8}\!=\!\num{94.4(2.9:2.7)}\%$, and $\num{97.3(0.4:0.4)}\%$ confinement to the $\left|\mathcal{C}^\pm_{1.8}\right\rangle$ subspace. Uncertainties in fidelity are 68.2\% bootstrap confidence intervals, while the cat size --- defined here as the displacement amplitude $|\alpha|$ --- carries a $\pm5\%$ $1\sigma$ uncertainty (see Methods\ref{met:bootstrap}). 

Using a linear drive, we apply an $X$-rotation to $\ecata{1.8}$ (\autoref{fig:cats}c), measuring an odd-parity cat state $\ocata{1.8}$ with
fidelity $\mathcal{F}^{-}_{1.8}\!=\!\num{74.8(1.9:1.9)}\%$ and
$\num{93.1(1.4:1.3)}\%$ confinement to the
$\left|\mathcal{C}^\pm_{1.8}\right\rangle$ subspace
(\autoref{fig:cats}d). This corresponds to a parity-inversion fidelity of
$\mathcal{F}_X^{\mathcal{C}}\!=\!\mathcal{F}^{-}_{1.8}/\mathcal{F}_{1.8}^{+}\!=\!\num{79.2(3.5:3.3)}\%$. The initial population in $|0\rangle$ sets an upper bound for the maximum possible even and odd cat-state fidelity of  $\num{96.5(3.5:3.5)}\%$, within the error of $\mathcal{F}^{+}_{1.8}$. The lower state fidelity of $|\mathcal{C}_{1.8}^-\rangle$ relative to $\ecata{1.8}$ is possibly a result of displacement noise in the linear drive, and imaging-related measurement error (see Methods\ref{met:TOF_sim}).

\subsubsection{Increasing cat size with painting}
\label{subsec:cat_size}

The ability to tune $|\alpha|$ is important for tailoring cat states to specific applications. In quantum metrology, larger cat states are advantageous for displacement sensing as the quantum Fisher information scales quadratically with $\abs{\alpha}$~\cite{fadel2025quantum}. Quantum computation, by contrast, benefits from comparatively smaller cats ($|\alpha| \approx 2$), whose coherent-state components are sufficiently orthogonal for high-fidelity logical gates~\cite{cochrane1999macroscopically} while being less susceptible to parity-mixing noise. In most Kerr-cat platforms, the nonlinearity is fixed, so the cat size can only be adjusted through the drive amplitude or detuning. While both provide some control, they cannot fully compensate for the limitation imposed by the fixed anharmonicity of the potential. In optical tweezers, however, this anharmonicity can be reduced via potential painting (\autoref{fig:kerr_figure}c), providing access to higher-lying motional states. 

We leverage this capability to prepare even-parity cat states across a broad range of sizes (\autoref{fig:cats}e), varying $\eta$ while keeping $\omega$ approximately fixed by increasing the trap depth. The cat size scales as $\abs{\alpha} \propto \sqrt{\tilde\varepsilon_q/\eta}$ (\autoref{eq:alpha}), where $\tilde\varepsilon_q \equiv \varepsilon_q - 4\delta/\omega$ is an effective drive amplitude. The measured values of $|\alpha|$ span $0.9$ to $2.7$, with the state fidelity decreasing from $\mathcal{F}_{0.9}^+=\num{95.4(1.8:1.5)}\%$ to $\mathcal{F}_{2.7}^+=\num{57.3(1.9:1.9)}\%$. We attribute this decline in part to tomographic reconstruction errors arising from limited imaging resolution, whose impact becomes significant for larger cats (see Methods\ref{met:TOF_sim}). The decline may also reflect the enhanced sensitivity of larger cats to  displacement noise --- a signature of the metrological advantage noted above --- which can induce parity mixing. 

For $\tilde{\varepsilon}_q \gtrsim 0.25$, the measured values of $|\alpha|$ fall below those predicted by the first-order model of \autoref{eq:alpha}. This discrepancy likely reflects a departure from the Kerr approximation at large $\tilde\varepsilon_q$, where higher-order nonlinearities become non-negligible (see Methods). While \autoref{eq:alpha} predicts that $|\alpha|$ increases monotonically with $\tilde \varepsilon_q$, the finite depth of the optical trap ultimately imposes a practical upper bound. For an approximately fixed drive and trap frequency, we show that a painted potential provides a significant increase in $|\alpha|$ compared to the native Gaussian trap, illustrating how painting can maximize cat size at a desired trap frequency.

\subsubsection{Cat states versus Fock states in tweezers}
\label{sec:cat_vs_fock}
While painting provides a method for exploring the large-$|\alpha|$ limit of a fixed-frequency trap, the same motional-control framework applies in the opposite limit of $|\alpha|\! \to\! 0$. In this regime, where coherent-state components of the cat overlap at the phase-space origin and converge to single Fock states, a non-adiabatic quadratic drive with pulse length  $T_{X}\gg1/|8K|$ enables isolation of the $\{\ket{0},\ket{2}\}$ subspace. Similarly, a purely linear drive with  pulse length  $T_{X} \gg1/|2K|$ can be used to address only the $\{\ket{0},\ket{1}\}$ subspace. 

\begin{figure}[t!]
\centering
\includegraphics[width=\columnwidth]{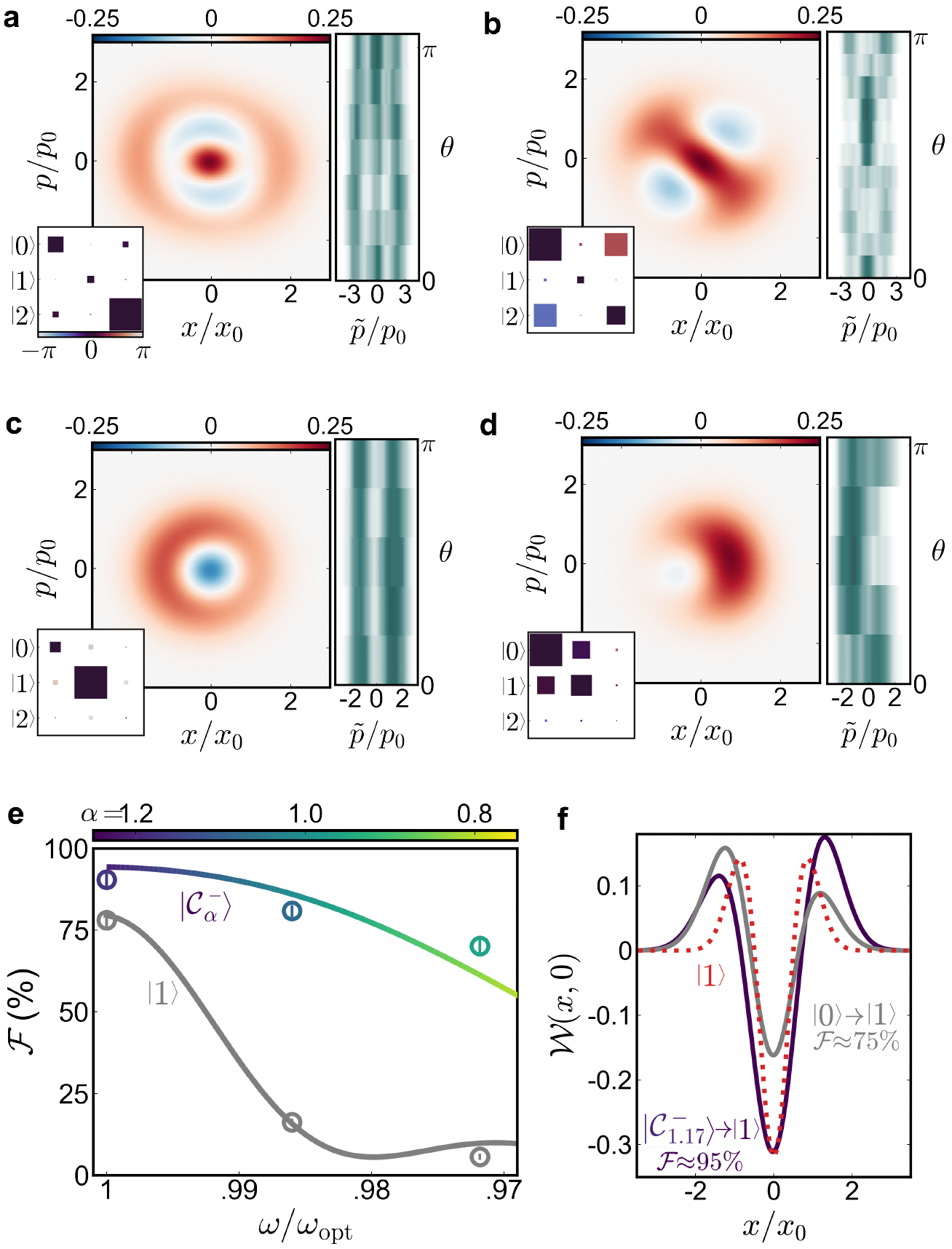} 
\caption{\textbf{Fock-state control with tweezer depth and position modulation}. \textbf{a,b,c,d,} Wigner functions, Hinton plots, and TOF quadrature measurements for motional Fock states $\ket{2}$ (a), $\propto\ket{0}+\ket{2}$  (b), $\ket{1}$ (c), and $\propto\ket{0}+\ket{1}$ (d), prepared via modulation of the tweezer depth (a,b) or position (c,d). Superposition states are prepared with $\sim1/2$ of the pulse amplitude and approximately the same pulse time used for $|2\rangle$ and $|1\rangle$. \textbf{e,}  State fidelity $\mathcal{F}$ of $\ket{1}$ and $|\mathcal{C}_{\alpha}^-\rangle$ as a function of fractional trap frequency $\omega/\omega_{\mathrm{opt}}$. Circles indicate experimental data while lines are simulated (see Methods\ref{met:trap_freq_and_fidelity}). \textbf{f,} Cross-sections of the Wigner functions $\mathcal{W}(x,p\!=\!0)$ for the state $\ket{1}$ in panel (c) prepared by applying a linear drive to $\ket{0}$ ($\ket{0}\rightarrow \ket{1}$, gray line) and via adiabatic mapping of an odd-cat state $|\mathcal{C}_{1.2}^-\rangle$ onto $|1\rangle$ ($|0\rangle\rightarrow |\mathcal{C}_{1.2}^+\rangle\rightarrow |\mathcal{C}_{1.2}^-\rangle \rightarrow \ket{1}$, purple line). As a guide to the eye, the red dotted line represents a pure $\ket{1}$. Error bars in (e) represent 68.2\% bootstrapped confidence intervals.}
\label{fig:fock}
\end{figure}

In an unpainted Gaussian potential, we independently apply these drives to $\ket{0}$, preparing $\ket{2}$ ($\ket{1}$) with state fidelity $\mathcal{F}^{\ket{2}}\!=\! \num{57.9(1.0:1.0)}\% \ (\mathcal{F}^{\ket{1}}\!=\! \num{75.2(1.4:1.5)}\%)$ [\autoref{fig:fock}a(c)], far exceeding $\mathcal{F}^{\ket{1}}$ obtained via tunnel-coupling in a $^{87}$Rb tweezer-based spatial double-well~\cite{brown2023time}. Accounting for preparation error in $\ket{0}$, this yields $\mathcal{F}^{\ket{2}}_X\!=\!\num{60.0(3.7:3.7)}\% \ (\mathcal{F}^{\ket{1}}_X\!=\!\num{78.0(3.8:3.9)}\%)$, where $\mathcal{F}^{\ket{n}}_X\!\equiv\!\mathcal{F}^{\ket{n}}\!/\!\mathcal{F}^{\ket{0}}$. Compared to the cat states in \autoref{fig:cats}, parity inversion of the Fock state required pulse durations ${\sim}7$ times longer to achieve comparable spectral selectivity (see Methods\ref{met:pulse_params}). For both excited Fock states, reducing the drive amplitude by $\sim1/2$ while keeping the pulse time approximately fixed generates the corresponding Fock-state superpositions (\autoref{fig:fock} b,d).

Excited Fock-state fidelities are constrained in part by imperfect three-dimensional ground-state cooling and shot-to-shot trap-frequency noise (see Methods). Cat states, by contrast, are remarkably insensitive to both. To quantify this robustness, we independently optimize the preparation of $\ocata{1.2}$ and $ \ket{1}$ at a trap frequency $\omega_{\mathrm{opt}}$, and perform tomography to extract $\mathcal{F}_{1.2}^{-}$ and $\mathcal{F}^{\ket{1}}$. We then introduce a controlled trap-frequency shift by slightly reducing the trap depth while applying the same pulse sequence optimized for $\omega_{\mathrm{opt}}$. As shown in \autoref{fig:fock}e, the cat state exhibits only a modest reduction in fidelity while the Fock state degrades rapidly with increasing fractional frequency shift $\omega/\omega_{\mathrm{opt}}$. 

This stark contrast between cat- and Fock-state sensitivity stems from a fundamental difference in how their parity rotations depend on trap frequency. For  Fock states, a small trap-frequency shift manifests as a drive detuning, $\omega_{\mathrm{opt}}-\omega=\omega_l-\omega\equiv\delta$, which for $|\delta|\ll\omega\varepsilon_l$, reduces $\mathcal{F}_X^{\ket{1}}$ proportional to $(2\delta/\omega\varepsilon_l)^2$. In our system, the estimated RMS fractional trap-frequency noise of \num{3.5e-3}  can account for a ${\sim}10\%$ reduction in $\mathcal{F}^{\ket{1}}$ (see Methods\ref{met:3D_pop_and_fidelity}). For cat states, the linear drive resonance is defined by the quadratic drive frequency rather than the trap frequency. Hence, the linear drive remains resonant by design, and a trap-frequency shift only detunes the quadratic drive from resonance. For $|\delta|\ll\omega\varepsilon_q$, a detuning shifts the nominal cat size as $2|\alpha| \delta/\omega\varepsilon_q$ (\autoref{eq:alpha}) while keeping the cat parity doublets degenerate, thereby changing the parity-inversion time and reducing $\mathcal{F}^{-}_{\alpha}$ proportional to $ (\pi\delta/\omega \varepsilon_q)^2$ (see Methods). 

This same protective mechanism also suppresses infidelity arising from motional excitations in orthogonal tweezer dimensions $(y,z)$. For the measured motional populations and estimated shot-to-shot trap-frequency noise in our experiment, these effects limit the fidelity of directly prepared excited Fock-states to $\mathcal{F}^{\ket{n}}\!\approx\!75\%$, while having negligible impact on cat-state fidelity $\mathcal{F}_{\alpha}^{\pm}$ (see Methods\ref{met:3D_pop_and_fidelity}). Leveraging this protective property of the cat encoding, we increase the fidelity of $\ket{1}$ from \num{75.2(1.4:1.3)}\% to \num{95.4(1.2:1.4)}\% by adiabatically mapping the odd-parity cat state $\ocata{1.2}$ onto $\ket{1}$ (see \autoref{fig:fock}f).  While alternative state-preparation techniques relying on adiabatic transfer or echo pulse sequences could offer similar improvements in $\mathcal{F}^{\ket{n}}$, they generally require longer pulse durations or more stringent parameter calibrations. 

The noise-protection benefits of the cat encoding are not unique to neutral atoms, but are also a defining feature of superconducting Kerr-cat qubits. In these systems, flux- and charge-noise are analogous to trap-frequency noise in optical tweezers~\cite{frattini2021three}. Other error mechanisms, however, differ substantially between the two platforms. Superconducting circuits are limited by spontaneous photon loss~\cite{putterman2025hardware,putterman2025preserving}, whereas optically trapped atoms experience recoil heating from off-resonant scattering of the tweezer light. While this process can become significant in deeper traps, it remains negligible ($\sim$0.1 phonons s$^{-1}$) for the shallow trap depths considered in this work.

\subsubsection{Outlook}
Looking forward, several technical improvements are within reach. Non-adiabatic shortcuts and optimal-control techniques could dramatically accelerate Kerr-based state preparation~\cite{kendell2024deterministic,grochowski2025quantum}, while single-shot readout may be realized through parity mapping onto spatially separated coherent states. Operating in deeper traps could provide additional benefits, including larger accessible cat sizes, faster gate operations, and coherent transfer of motional information to internal states. However, these theoretical advantages must be balanced against increased recoil heating and other potential decoherence mechanisms.

By establishing motional Kerr-cat states as a new resource within the optical-tweezer platform, our work opens a path toward bosonic quantum information processing, quantum simulation, and quantum-enhanced sensing with reconfigurable neutral-atom arrays. More broadly, our work demonstrates that dynamic optical tweezers provide a universal motional-control framework that elevates motion from an experimental constraint to a robust degree of freedom, complementing the well-established internal-state toolbox of neutral atoms and expanding the range of quantum resources in optically trapped matter.
\bibliographystyle{ieeetr}
\bibliography{references}

@article{Kasevich1991,
author = {Kasevich, Mark and Chu, Steven},
title = {Atomic Interferometry Using Stimulated Raman Transitions},
journal = {Physical Review Letters},
volume = {67},
number = {2},
pages = {181--184},
year = {1991},
doi = {10.1103/PhysRevLett.67.181}
}

@article{Lo2015,
author = {Lo, H.-Y. and Kienzler, D. and Keitch, B. and de Clercq, L. and Leupold, F. and Lindenfelser, F. and Marinelli, M. and Negnevitsky, V. and Home, J. P.},
title = {Spin--Motion Entanglement and State Diagnosis with Squeezed Oscillator Wavepackets},
journal = {Nature},
volume = {521},
pages = {336--339},
year = {2015},
doi = {10.1038/nature14458}
}

@article{Brune1996,
author = {Brune, M. and Hagley, E. and Dreyer, J. and Maitre, X. and Maali, A. and Wunderlich, C. and Raimond, J. M. and Haroche, S.},
title = {Observing the Progressive Decoherence of the ``Meter'' in a Quantum Measurement},
journal = {Physical Review Letters},
volume = {77},
pages = {4887--4890},
year = {1996},
doi = {10.1103/PhysRevLett.77.4887}
}

@article{Friedman2000,
author = {Friedman, J. R. and Patel, V. and Chen, W. and Tolpygo, S. K. and Lukens, J. E.},
title = {Quantum Superposition of Distinct Macroscopic States},
journal = {Nature},
volume = {406},
pages = {43--46},
year = {2000},
doi = {10.1038/35017505}
}

@article{Vlastakis2013,
author = {Vlastakis, B. and Kirchmair, G. and Leghtas, Z. and Nigg, S. E. and Frunzio, L. and Girvin, S. M. and Mirrahimi, M. and Devoret, M. H. and Schoelkopf, R. J.},
title = {Deterministically Encoding Quantum Information Using 100-Photon {Schr{\"o}dinger} Cat States},
journal = {Science},
volume = {342},
number = {6158},
pages = {607--610},
year = {2013},
doi = {10.1126/science.1243289}
}

@article{Greve2022Entanglement,
   author = {Greve, Graham P. and Luo, Chengyi and Wu, Baochen and Thompson, James K.},
   title = {Entanglement-enhanced matter-wave interferometry in a high-finesse cavity},
   journal = {Nature},
   volume = {610},
   number = {7932},
   pages = {472-477},
   ISSN = {1476-4687},
   DOI = {10.1038/s41586-022-05197-9},
   year = {2022},
   type = {Journal Article}
}

@article{Wolf2019Motional,
   author = {Wolf, Fabian and Shi, Chunyan and Heip, Jan C. and Gessner, Manuel and Pezzè, Luca and Smerzi, Augusto and Schulte, Marius and Hammerer, Klemens and Schmidt, Piet O.},
   title = {Motional {Fock} states for quantum-enhanced amplitude and phase measurements with trapped ions},
   journal = {Nature Communications},
   volume = {10},
   number = {1},
   pages = {2929},
   ISSN = {2041-1723},
   DOI = {10.1038/s41467-019-10576-4},
   year = {2019},
   type = {Journal Article}
}

@article{mccormick2019quantum,
author = {McCormick, Katherine C. and Keller, Jonas and Burd, Shaun C. and Wineland, David J. and Wilson, Andrew C. and Leibfried, Dietrich},
title = {Quantum-enhanced sensing of a mechanical oscillator},
journal = {Nature},
volume = {572},
pages = {86--90},
year = {2019},
doi = {10.1038/s41586-019-1421-y}
}

@article{Cassens2025Entanglement,
  title = {Entanglement-Enhanced Atomic Gravimeter},
  author = {Cassens, Christophe and Meyer-Hoppe, Bernd and Rasel, Ernst and Klempt, Carsten},
  journal = {Physical Review X},
  volume = {15},
  issue = {1},
  pages = {011029},
  numpages = {7},
  year = {2025},
  month = {Feb},
  publisher = {American Physical Society},
  doi = {10.1103/PhysRevX.15.011029},
}

@article{grimm2020stabilization,
  title={Stabilization and operation of a {Kerr-cat} qubit},
  author={Grimm, Alexander and Frattini, Nicholas E and Puri, Shruti and Mundhada, Shantanu O and Touzard, Steven and Mirrahimi, Mazyar and Girvin, Steven M and Shankar, Shyam and Devoret, Michel H},
  journal={Nature},
  volume={584},
  number={7820},
  pages={205--209},
  year={2020},
  publisher={Nature Publishing Group UK London}
}

@article{leghtas2015confining,
  title={Confining the state of light to a quantum manifold by engineered two-photon loss},
  author={Leghtas, Zaki and Touzard, Steven and Pop, Ioan M and Kou, Angela and Vlastakis, Brian and Petrenko, Andrei and Sliwa, Katrina M and Narla, Anirudh and Shankar, Shyam and Hatridge, Michael J and others},
  journal={Science},
  volume={347},
  number={6224},
  pages={853--857},
  year={2015},
  publisher={American Association for the Advancement of Science}
}

@article{shaw2025erasure,
  title={Erasure cooling, control, and hyperentanglement of motion in optical tweezers},
  author={Shaw, Adam L and Scholl, Pascal and Finkelstein, Ran and Tsai, Richard Bing-Shiun and Choi, Joonhee and Endres, Manuel},
  journal={Science},
  volume={388},
  number={6749},
  pages={845--849},
  year={2025},
  publisher={American Association for the Advancement of Science}
}

@article{lienhard2025generation,
  title={Generation of Motional Squeezed States for Neutral Atoms in Optical Tweezers},
  author={Lienhard, Vincent and Martin, Romain and Chew, Yuki Torii and Tomita, Takafumi and Ohmori, Kenji and de L{\'e}s{\'e}leuc, Sylvain},
  journal={Physical Review Letters},
  volume={135},
  number={25},
  pages={253404},
  year={2025},
  publisher={APS}
}

@article{hartke2022quantum,
  title={Quantum register of fermion pairs},
  author={Hartke, Thomas and Oreg, Botond and Jia, Ningyuan and Zwierlein, Martin},
  journal={Nature},
  volume={601},
  number={7894},
  pages={537--541},
  year={2022},
  publisher={Nature Publishing Group UK London}
}

@article{lunt2024realization,
  title={Realization of a {Laughlin} state of two rapidly rotating fermions},
  author={Lunt, Philipp and Hill, Paul and Reiter, Johannes and Preiss, Philipp M and Ga{\l}ka, Maciej and Jochim, Selim},
  journal={Physical Review Letters},
  volume={133},
  number={25},
  pages={253401},
  year={2024},
  publisher={APS}
}

@article{bohnmann2025bosonic,
  title={Bosonic quantum error correction with neutral atoms in optical dipole traps},
  author={Bohnmann, Leon H and Locher, David F and Zeiher, Johannes and M{\"u}ller, Markus},
  journal={Physical Review A},
  volume={111},
  number={2},
  pages={022432},
  year={2025},
  publisher={APS}
}

@article{murmann2015two,
  title={Two fermions in a double well: Exploring a fundamental building block of the {Hubbard} model},
  author={Murmann, Simon and Bergschneider, Andrea and Klinkhamer, Vincent M and Z{\"u}rn, Gerhard and Lompe, Thomas and Jochim, Selim},
  journal={Physical Review Letters},
  volume={114},
  number={8},
  pages={080402},
  year={2015},
  publisher={APS}
}

@article{sandholzer2022floquet,
  title={Floquet engineering of individual band gaps in an optical lattice using a two-tone drive},
  author={Sandholzer, Kilian and Walter, Anne-Sophie and Minguzzi, Joaqu{\'\i}n and Zhu, Zijie and Viebahn, Konrad and Esslinger, Tilman},
  journal={Physical Review Research},
  volume={4},
  number={1},
  pages={013056},
  year={2022},
  publisher={APS}
}

@article{brown2023time,
  title={Time-of-flight quantum tomography of an atom in an optical tweezer},
  author={Brown, MO and Muleady, SR and Dworschack, WJ and Lewis-Swan, RJ and Rey, AM and Romero-Isart, O and Regal, CA},
  journal={Nature Physics},
  volume={19},
  number={4},
  pages={569--573},
  year={2023},
  publisher={Nature Publishing Group UK London}
}

@phdthesis{serwane2011deterministic,
  title={Deterministic preparation of a tunable few-fermion system},
  author={Serwane, Friedhelm},
  year={2011},
  school={Ruperto-Carola-University of Heidelberg}
}

@article{sych2012informational,
  title={Informational completeness of continuous-variable measurements},
  author={Sych, D and {\v{R}}eh{\'a}{\v{c}}ek, J and Hradil, Z and Leuchs, G and S{\'a}nchez-Soto, LL},
  journal={Physical Review A—Atomic, Molecular, and Optical Physics},
  volume={86},
  number={5},
  pages={052123},
  year={2012},
  publisher={APS}
}

@article{monroe1996schrodinger,
  title={A {“Schr{\"o}dinger cat”} superposition state of an atom},
  author={Monroe, Christopher and Meekhof, Dawn M and King, Brian E and Wineland, David J},
  journal={Science},
  volume={272},
  number={5265},
  pages={1131--1136},
  year={1996},
  publisher={American Association for the Advancement of Science}
}

@article{facon2016sensitive,
  title={A sensitive electrometer based on a {Rydberg} atom in a {Schr{\"o}dinger-cat} state},
  author={Facon, Adrien and Dietsche, Eva-Katharina and Grosso, Dorian and Haroche, Serge and Raimond, Jean-Michel and Brune, Michel and Gleyzes, S{\'e}bastien},
  journal={Nature},
  volume={535},
  number={7611},
  pages={262--265},
  year={2016},
  publisher={Nature Publishing Group UK London}
}

@article{cao2024multi,
  title={Multi-qubit gates and {Schr{\"o}dinger} cat states in an optical clock},
  author={Cao, Alec and Eckner, William J and Lukin Yelin, Theodor and Young, Aaron W and Jandura, Sven and Yan, Lingfeng and Kim, Kyungtae and Pupillo, Guido and Ye, Jun and Darkwah Oppong, Nelson and others},
  journal={Nature},
  volume={634},
  number={8033},
  pages={315--320},
  year={2024},
  publisher={Nature Publishing Group UK London}
}

@article{yang2025minute,
  title={Minute-scale {Schr{\"o}dinger-cat} state of spin-5/2 atoms},
  author={Yang, YA and Luo, W-T and Zhang, J-L and Wang, S-Z and Zou, Chang-Ling and Xia, T and Lu, Z-T},
  journal={Nature Photonics},
  volume={19},
  number={1},
  pages={89--94},
  year={2025},
  publisher={Nature Publishing Group UK London}
}

@article{steffen2012digital,
  title={Digital atom interferometer with single particle control on a discretized space-time geometry},
  author={Steffen, Andreas and Alberti, Andrea and Alt, Wolfgang and Belmechri, Noomen and Hild, Sebastian and Karski, Micha{\l} and Widera, Artur and Meschede, Dieter},
  journal={Proceedings of the National Academy of Sciences},
  volume={109},
  number={25},
  pages={9770--9774},
  year={2012},
  publisher={National Academy of Sciences}
}

@article{parazzoli2012observation,
  title={Observation of free-space single-atom matter wave interference},
  author={Parazzoli, Lambert Paul and Hankin, Aaron M and Biedermann, Grant W},
  journal={Physical Review Letters},
  volume={109},
  number={23},
  pages={230401},
  year={2012},
  publisher={APS}
}

@article{gonzalez2023fermionic,
  title={Fermionic quantum processing with programmable neutral atom arrays},
  author={Gonz{\'a}lez-Cuadra, Daniel and Bluvstein, Dolev and Kalinowski, Marcin and Kaubruegger, Raphael and Maskara, Nishad and Naldesi, Piero and Zache, Torsten V and Kaufman, Adam M and Lukin, Mikhail D and Pichler, Hannes and others},
  journal={Proceedings of the National Academy of Sciences},
  volume={120},
  number={35},
  pages={e2304294120},
  year={2023},
  publisher={National Academy of Sciences}
}

@article{grochowski2025quantum,
  title={Quantum control of continuous systems via nonharmonic potential modulation},
  author={Grochowski, Piotr T and Pichler, Hannes and Regal, Cindy A and Romero-Isart, Oriol},
  journal={Quantum},
  volume={9},
  pages={1824},
  year={2025},
  publisher={Verein zur F{\"o}rderung des Open Access Publizierens in den Quantenwissenschaften}
}

@article{pedalino2026probing,
  title={Probing quantum mechanics with nanoparticle matter-wave interferometry},
  author={Pedalino, Sebastian and Ram{\'\i}rez-Galindo, Bruno E and Ferstl, Richard and Hornberger, Klaus and Arndt, Markus and Gerlich, Stefan},
  journal={Nature},
  volume={649},
  number={8098},
  pages={866--870},
  year={2026},
  publisher={Nature Publishing Group UK London}
}

@article{fein2019quantum,
  title={Quantum superposition of molecules beyond {25 kDa}},
  author={Fein, Yaakov Y and Geyer, Philipp and Zwick, Patrick and Kia{\l}ka, Filip and Pedalino, Sebastian and Mayor, Marcel and Gerlich, Stefan and Arndt, Markus},
  journal={Nature Physics},
  volume={15},
  number={12},
  pages={1242--1245},
  year={2019},
  publisher={Nature Publishing Group UK London}
}

@article{johnson2017ultrafast,
  title={Ultrafast creation of large {Schr{\"o}dinger} cat states of an atom},
  author={Johnson, KG and Wong-Campos, JD and Neyenhuis, B and Mizrahi, J and Monroe, C},
  journal={Nature Communications},
  volume={8},
  number={1},
  pages={697},
  year={2017},
  publisher={Nature Publishing Group UK London}
}

@article{putterman2025hardware,
  title={Hardware-efficient quantum error correction via concatenated bosonic qubits},
  author={Putterman, Harald and Noh, Kyungjoo and Hann, Connor T and MacCabe, Gregory S and Aghaeimeibodi, Shahriar and Patel, Rishi N and Lee, Menyoung and Jones, William M and Moradinejad, Hesam and Rodriguez, Roberto and others},
  journal={Nature},
  volume={638},
  number={8052},
  pages={927--934},
  year={2025},
  publisher={Nature Publishing Group UK London}
}

@article{fadel2025quantum,
  title={Quantum metrology with a continuous-variable system},
  author={Fadel, Matteo and Roux, Noah and Gessner, Manuel},
  journal={Reports on Progress in Physics},
  volume={88},
  number={10},
  pages={106001},
  year={2025},
  publisher={IOP Publishing}
}

@article{bild2023schrodinger,
  title={{Schr{\"o}dinger} cat states of a 16-microgram mechanical oscillator},
  author={Bild, Marius and Fadel, Matteo and Yang, Yu and Von L{\"u}pke, Uwe and Martin, Phillip and Bruno, Alessandro and Chu, Yiwen},
  journal={Science},
  volume={380},
  number={6642},
  pages={274--278},
  year={2023},
  publisher={American Association for the Advancement of Science}
}

@article{putterman2025preserving,
  title={Preserving phase coherence and linearity in cat qubits with exponential bit-flip suppression},
  author={Putterman, Harald and Noh, Kyungjoo and Patel, Rishi N and Peairs, Gregory A and MacCabe, Gregory S and Lee, Menyoung and Aghaeimeibodi, Shahriar and Hann, Connor T and Jarrige, Ignace and Marcaud, Guillaume and others},
  journal={Physical Review X},
  volume={15},
  number={1},
  pages={011070},
  year={2025},
  publisher={APS}
}

@article{aliferis2008fault,
  title={Fault-tolerant quantum computation against biased noise},
  author={Aliferis, Panos and Preskill, John},
  journal={Physical Review A—Atomic, Molecular, and Optical Physics},
  volume={78},
  number={5},
  pages={052331},
  year={2008},
  publisher={APS}
}

@article{kaufman2012cooling,
  title={Cooling a single atom in an optical tweezer to its quantum ground state},
  author={Kaufman, Adam M and Lester, Brian J and Regal, Cindy A},
  journal={Physical Review X},
  volume={2},
  number={4},
  pages={041014},
  year={2012},
  publisher={APS}
}

@article{fluhmann2019encoding,
  title={Encoding a qubit in a trapped-ion mechanical oscillator},
  author={Fl{\"u}hmann, Christa and Nguyen, Thanh Long and Marinelli, Matteo and Negnevitsky, Vlad and Mehta, Karan and Home, Jonathan P},
  journal={Nature},
  volume={566},
  number={7745},
  pages={513--517},
  year={2019},
  publisher={Nature Publishing Group UK London}
}

@article{yang2024mechanical,
  title={A mechanical qubit},
  author={Yang, Yu and Kladari{\'c}, Igor and Drimmer, Maxwell and von L{\"u}pke, Uwe and Lenterman, Daan and Bus, Joost and Marti, Stefano and Fadel, Matteo and Chu, Yiwen},
  journal={Science},
  volume={386},
  number={6723},
  pages={783--788},
  year={2024},
  publisher={American Association for the Advancement of Science}
}

@article{puri2019cat,
  title = {Stabilized Cat in a Driven Nonlinear Cavity: A Fault-Tolerant Error Syndrome Detector},
  author = {Puri, Shruti and Grimm, Alexander and Campagne-Ibarcq, Philippe and Eickbusch, Alec and Noh, Kyungjoo and Roberts, Gabrielle and Jiang, Liang and Mirrahimi, Mazyar and Devoret, Michel H. and Girvin, S. M.},
  journal = {Physical Review X},
  volume = {9},
  issue = {4},
  pages = {041009},
  numpages = {29},
  year = {2019},
  month = {Oct},
  publisher = {American Physical Society}
}

@article{levine2018high,
  title={High-fidelity control and entanglement of {Rydberg-atom} qubits},
  author={Levine, Harry and Keesling, Alexander and Omran, Ahmed and Bernien, Hannes and Schwartz, Sylvain and Zibrov, Alexander S and Endres, Manuel and Greiner, Markus and Vuleti{\'c}, Vladan and Lukin, Mikhail D},
  journal={Physical Review Letters},
  volume={121},
  number={12},
  pages={123603},
  year={2018},
  publisher={APS}
}

@article{qing2026quantum,
  title={Quantum benchmarking of high-fidelity noise-biased operations on a detuned {Kerr-cat} qubit},
  author={Qing, Bingcheng and Hajr, Ahmed and Wang, Ke and Koolstra, Gerwin and Nguyen, Long B and Hines, Jordan and Huang, Irwin and Bhandari, Bibek and Chen, Larry and Kang, Ziqi and others},
  journal={Proceedings of the National Academy of Sciences},
  volume={123},
  number={5},
  pages={e2520479123},
  year={2026},
  publisher={National Academy of Sciences}
}

@article{kaufman2021quantum,
  title={Quantum science with optical tweezer arrays of ultracold atoms and molecules},
  author={Kaufman, Adam M and Ni, Kang-Kuen},
  journal={Nature Physics},
  volume={17},
  number={12},
  pages={1324--1333},
  year={2021},
  publisher={Nature Publishing Group UK London}
}

@article{spar2022realization,
  title={Realization of a {Fermi-Hubbard} optical tweezer array},
  author={Spar, Benjamin M and Guardado-Sanchez, Elmer and Chi, Sungjae and Yan, Zoe Z and Bakr, Waseem S},
  journal={Physical Review Letters},
  volume={128},
  number={22},
  pages={223202},
  year={2022},
  publisher={APS}
}

@article{henriet2020quantum,
  title={Quantum computing with neutral atoms},
  author={Henriet, Lo{\"\i}c and Beguin, Lucas and Signoles, Adrien and Lahaye, Thierry and Browaeys, Antoine and Reymond, Georges-Olivier and Jurczak, Christophe},
  journal={Quantum},
  volume={4},
  pages={327},
  year={2020},
  publisher={Verein zur F{\"o}rderung des Open Access Publizierens in den Quantenwissenschaften}
}

@article{gross2017quantum,
  title={Quantum simulations with ultracold atoms in optical lattices},
  author={Gross, Christian and Bloch, Immanuel},
  journal={Science},
  volume={357},
  number={6355},
  pages={995--1001},
  year={2017},
  publisher={American Association for the Advancement of Science}
}

@article{wirth2011evidence,
  title={Evidence for orbital superfluidity in the P-band of a bipartite optical square lattice},
  author={Wirth, Georg and {\"O}lschl{\"a}ger, Matthias and Hemmerich, Andreas},
  journal={Nature Physics},
  volume={7},
  number={2},
  pages={147--153},
  year={2011},
  publisher={Nature Publishing Group UK London}
}

@article{zheng2026quantum,
  title={Quantum-Enhanced Dark Matter Search Using Cat States},
  author={Zheng, Pan and Cai, Yanyan and Xu, Bin and Wen, Shengcheng and Zhang, Libo and Ni, Zhongchu and Mai, Jiasheng and Zeng, Yanjie and Lin, Lin and Hu, Ling and others},
  journal={Physical Review Letters},
  volume={136},
  number={17},
  pages={171002},
  year={2026},
  publisher={APS}
}

@article{cortinas2024towards,
  title={Towards the generation of mechanical {Kerr-cats}: awakening the perturbative quantum Moyal corrections to classical motion},
  author={Cortinas, Rodrigo G},
  journal={New Journal of Physics},
  volume={26},
  number={2},
  pages={023022},
  year={2024},
  publisher={IOP Publishing}
}

@article{kendell2024deterministic,
  title={Deterministic generation of highly squeezed {GKP} states in ultracold atoms},
  author={Kendell, Harry CP and Ferranti, Giacomo and Weidner, Carrie A},
  journal={APL Quantum},
  volume={1},
  number={2},
  year={2024},
  publisher={AIP Publishing}
}

@article{milburn1986quantum,
  title={Quantum and classical {Liouville} dynamics of the anharmonic oscillator},
  author={Milburn, Gerard J},
  journal={Physical Review A},
  volume={33},
  number={1},
  pages={674--685},
  year={1986},
  publisher={American Physical Society}
}

@article{yurke1986generating,
  title={Generating quantum mechanical superpositions of macroscopically distinguishable states via amplitude dispersion},
  author={Yurke, Bernard and Stoler, David},
  journal={Physical Review Letters},
  volume={57},
  number={1},
  pages={13},
  year={1986},
  publisher={APS}
}

@article{kaufman2015entangling,
  title={Entangling two transportable neutral atoms via local spin exchange},
  author={Kaufman, AM and Lester, BJ and Foss-Feig, M and Wall, ML and Rey, AM and Regal, CA},
  journal={Nature},
  volume={527},
  number={7577},
  pages={208--211},
  year={2015},
  publisher={Nature Publishing Group UK London}
}

@article{omran2019generation,
  title={Generation and manipulation of {Schr{\"o}dinger} cat states in {Rydberg} atom arrays},
  author={Omran, Ahmed and Levine, Harry and Keesling, Alexander and Semeghini, Giulia and Wang, Tout T and Ebadi, Sepehr and Bernien, Hannes and Zibrov, Alexander S and Pichler, Hannes and Choi, Soonwon and others},
  journal={Science},
  volume={365},
  number={6453},
  pages={570--574},
  year={2019},
  publisher={American Association for the Advancement of Science}
}

@article{delic2020cooling,
  title={Cooling of a levitated nanoparticle to the motional quantum ground state},
  author={Deli{\'c}, Uro{\v{s}} and Reisenbauer, Manuel and Dare, Kahan and Grass, David and Vuleti{\'c}, Vladan and Kiesel, Nikolai and Aspelmeyer, Markus},
  journal={Science},
  volume={367},
  number={6480},
  pages={892--895},
  year={2020},
  publisher={American Association for the Advancement of Science}
}

@article{bergschneider2019experimental,
  title={Experimental characterization of two-particle entanglement through position and momentum correlations},
  author={Bergschneider, Andrea and Klinkhamer, Vincent M and Becher, Jan Hendrik and Klemt, Ralf and Palm, Lukas and Z{\"u}rn, Gerhard and Jochim, Selim and Preiss, Philipp M},
  journal={Nature Physics},
  volume={15},
  number={7},
  pages={640--644},
  year={2019},
  publisher={Nature Publishing Group UK London}
}

@article{toscano2006sub,
  title={{Sub-Planck} phase-space structures and {Heisenberg-limited} measurements},
  author={Toscano, Fabricio and Dalvit, Diego AR and Davidovich, Luiz and Zurek, Wojciech H},
  journal={Physical Review A—Atomic, Molecular, and Optical Physics},
  volume={73},
  number={2},
  pages={023803},
  year={2006},
  publisher={APS}
}

@article{eckardt2017colloquium,
  title={Colloquium: Atomic quantum gases in periodically driven optical lattices},
  author={Eckardt, Andr{\'e}},
  journal={Reviews of Modern Physics},
  volume={89},
  number={1},
  pages={011004},
  year={2017},
  publisher={APS}
}

@article{henderson2009experimental,
  title={Experimental demonstration of painting arbitrary and dynamic potentials for {Bose--Einstein} condensates},
  author={Henderson, Kevin and Ryu, Changhyun and MacCormick, Calum and Boshier, MG},
  journal={New Journal of Physics},
  volume={11},
  number={4},
  pages={043030},
  year={2009}
}

@article{zurek2001sub,
  title={{Sub-Planck} structure in phase space and its relevance for quantum decoherence},
  author={Zurek, Wojciech Hubert},
  journal={Nature},
  volume={412},
  number={6848},
  pages={712--717},
  year={2001},
  publisher={Nature Publishing Group UK London}
}

@phdthesis{brown2022light,
  title={Light-Assisted Collisions and Quantum State Tomography of Single-Atom Motion in Optical Tweezers},
  author={Brown, Mark O},
  year={2022},
  school={University of Colorado at Boulder}
}

@phdthesis{frattini2021three,
  title={Three-wave mixing in superconducting circuits: stabilizing cats with SNAILs},
  author={Frattini, Nicholas E},
  year={2021},
  school={Yale University}
}

@article{campagne2020quantum,
  title={Quantum error correction of a qubit encoded in grid states of an oscillator},
  author={Campagne-Ibarcq, Philippe and Eickbusch, Alec and Touzard, Steven and Zalys-Geller, Evan and Frattini, Nicholas E and Sivak, Volodymyr V and Reinhold, Philip and Puri, Shruti and Shankar, Shyam and Schoelkopf, Robert J and others},
  journal={Nature},
  volume={584},
  number={7821},
  pages={368--372},
  year={2020},
  publisher={Nature Publishing Group UK London}
}

@article{cochrane1999macroscopically,
  title={Macroscopically distinct quantum-superposition states as a bosonic code for amplitude damping},
  author={Cochrane, Paul T and Milburn, Gerard J and Munro, William J},
  journal={Physical Review A},
  volume={59},
  number={4},
  pages={2631},
  year={1999},
  publisher={APS}
}

\section{Methods}\label{sec:methods}

\subsection{Potential painting in optical tweezers }\label{met:painted_tweezer_theory}

\begin{figure*}[t!]
    \centering
    \includegraphics[width=0.7\textwidth]{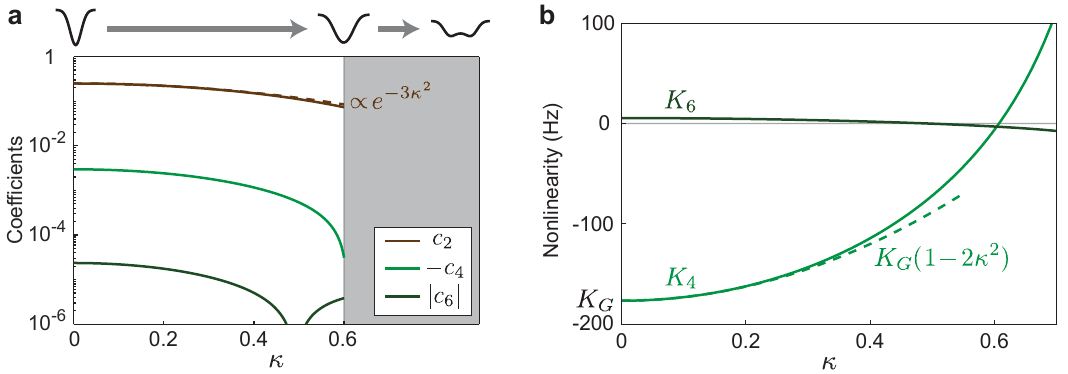}
    \caption{\textbf{Tunable Kerr nonlinearity in the painted trap.}
    \textbf{a,} Coefficients $c_2,c_4,c_6$ of the painted potential
    (\autoref{eq:painted_potential_exp}) as functions of the painting
    amplitude $\kappa$. As $\kappa$ increases, the original Gaussian
    potential becomes nearly harmonic around $\kappa\approx0.6$ and evolves
    toward a double-well potential for larger $\kappa$ (shaded region).
    \textbf{b,} Nonlinearities $K_4$ and $K_6$ derived from the coefficients
    in (a). Here, the trap frequency is held constant by increasing the trap depth.}
    \label{fig:painting_coeffs}
\end{figure*}

Following the expressions defined in section \textit{\nameref{subsec:kerr}}, we use the painting technique~\cite{henderson2009experimental} to generate a time-averaged (painted) potential along the radial $x$-axis of an optical tweezer with a Gaussian intensity profile. For painting frequency $\omega_p$ and amplitude $\kappa$, the painted potential is given by 
\begin{equation}
\label{eq:time_averaged_potential}
V(x)\!=\!\frac{-{V}_G}{2\pi}\!\int_0^{2\pi} \!\exp(-2(x-\kappa W\sin(\omega_p t))^2/W^2)\mathrm{d}(\omega_pt),   
\end{equation}
where ${V}_G$ is the Gaussian trap depth ($\kappa=0$) with harmonic trap frequency $\omega_{G} \equiv \sqrt{{4{V}_G}/{mW^2}}$, oscillator length $x_{G} = \sqrt{\hbar/(2m\omega_{G})}$, and waist $W$. The validity of this painted potential generally requires that our painting frequency be sufficiently fast, i.e. $\omega_p \gg \omega_{G}$. The potential can be expanded as
\begin{equation}
\label{eq:painted_potential_exp}
    \frac{V( x)}{\hbar\omega_{G}} = c_2 \left(\frac{x}{x_{G}}\right)^2 + c_4 \left(\frac{x}{x_{G}}\right)^4 + c_6 \left(\frac{x}{x_{G}}\right)^6 + \cdots.
\end{equation}
The coefficients $\{c_2,c_4,c_6\}$ are plotted as functions of $\kappa$ in \autoref{fig:painting_coeffs}(a).

As $\kappa$ increases, the potential shifts from a Gaussian to harmonic, quartic, and eventually double-well form. For $0\le\kappa\lesssim0.6$, the quadratic coefficient approximately follows $c_2 \approx e^{-3\kappa^2}/4$, and so the painted harmonic trap frequency is given by $\omega_{\kappa} \approx \omega_{G} e^{-3\kappa^2/2}$ with corresponding length scale $x_{\kappa} = \sqrt{\hbar/(2m\omega_{\kappa)}}$. The painted potential may then be rewritten as
\begin{equation}\label{eq:H_painted}
    \frac{V(x)}{\hbar} \!\approx \!\frac{\omega_{\kappa}}{4}\left(\frac{x}{x_{\kappa}}\right)^2 +\frac{K_4}{6}\left(\frac{x}{x_{\kappa}}\right)^4 \!+ \frac{K_6}{20}\left(\frac{x}{x_{\kappa}}\right)^6 \!+ \cdots
\end{equation}
with $K_4$ and $K_6$ shown in \autoref{fig:painting_coeffs}b for fixed $\omega_\kappa$. 

As shown in the next section, in the regime of weak anharmonicity, the Kerr nonlinearity $K$ may be expressed as $K \approx K_4$. For $\kappa\lesssim 0.5$, this scales as $K \approx K_4 \approx K_G(1-2\kappa^2)$, with $K_G\equiv-3\hbar/(4mW^2)=2\pi\times(-177)$ Hz for a Gaussian trap with a typical beam waist $W=\SI{0.7}{\micro m}$. $K$ can therefore be tuned continuously from $K_G$ to the desired value by varying the parameter $\kappa$. We neglect the sixth-order nonlinearity $|K_6|$ as it remains much smaller than $|K_4|$ in the experimentally relevant regime ($2K/2\pi \lesssim \SI{-100}{Hz}$).

\subsection{Kerr-Cat Hamiltonian}
\label{met:derivation}
The painted potential (\autoref{eq:H_painted}) with trap-depth modulation (\autoref{eq:quadratic_drive}) gives the Hamiltonian, which we now write in operator form as
\begin{equation}
\Hat{H}= \hat{H}_{0} + \Big{(}1+\varepsilon_q \cos(\omega_q t)\Big{)}{V}({\hat x}),
\end{equation}
where $\hat{H}_0 = \hat{p}^2/(2m)$ is the kinetic term for momentum operator $\hat{p}$, $\hat{x}$ is the position operator, and $\varepsilon_q$ ($\omega_q$) is the quadratic drive amplitude (frequency). We define the ladder operators $\hat{a}$ and $\hat{a}^{\dagger}$,
expressed in the harmonic basis defined by some frequency $\omega$, i.e.  $\hat{x} = x_0(\hat{a}^\dagger + \hat{a})$ and $\hat{p} = i p_0(\hat{a}^\dagger - \hat{a})$, with corresponding 
harmonic length $x_0 = \sqrt{\hbar/(2m\omega)}$ and momentum scale $p_0 = \sqrt{m\hbar\omega/2}$. We refer to the time-independent portion of the modulated potential as the \emph{static potential}, and the term proportional to $\varepsilon_q \cos(\omega_q t)$ as the \emph{drive}.

To isolate the slowest moving dynamics, we consider the interaction picture rotating at  $\omega_q/2 \approx \omega$. Our ladder operators then evolve as $\hat{a}e^{-i\omega_q t/2}$ and  $\hat{a}^{\dagger}e^{i\omega_q t/2}$. Expanding ${V}({\hat{x}})$ in terms of these ladder operators generates terms rotating at different harmonics of $\omega_q/2$. Within the rotating-wave approximation, we neglect all rapidly oscillating contributions and retain only time-independent terms in the interaction frame. Consequently, number-conserving operators diagonal in the Fock basis, such as $\hat{a}^{\dagger}\hat{a}$ and $\hat{a}^{\dagger2}\hat{a}^{2}$, contribute only through the static potential, while the modulated drive contributes only through terms proportional to $(\hat{a}^{\dagger})^{2}$ and $\hat{a}^{2}$.

To quartic order in \autoref{eq:H_painted}, the resulting diagonal portion of the Hamiltonian is exactly given by
\begin{align}
\hat{H}_{\rm diag.}/\hbar
\!=\!
-\delta\hat{a}^\dagger\hat{a}
+K\hat{a}^{\dagger2}\hat{a}^2,
\end{align}
for $\delta = \omega_q/2 - \omega$, where we have fixed $\omega = \omega_\kappa/\sqrt{1-2\eta}$ to correspond to the $\omega_{01}$ transition frequency and set $K = K_4(1-2\eta)$ to correspond to the Kerr nonlinearity. We have also introduced the relative anharmonicity $\eta \equiv 2K/\omega$. 

In terms of the painted harmonic trap frequency $\omega_{\kappa}$ and the quartic nonlinearity $K_4$, we introduce $\eta_{4,\kappa} \equiv 2K_4/\omega_{\kappa}$, and the relative anharmonicity is then self-consistently determined via $\eta (1-2\eta)^{-3/2} = \eta_{4,\kappa}$. In the limit $|\eta_{4,\kappa}| \ll 1$, we have $\eta \approx \eta_{4,\kappa} - 3\eta_{4,\kappa}^2 + \cdots$. Thus, to leading order in $\eta_{4,\kappa}$, we obtain $\omega \approx \omega_{\kappa} + 2K_4$ and $K \approx K_4(1 - \eta_{4,\kappa})$.

We also have the addition of an off-diagonal anti-squeezing term, given by 
\begin{equation}\label{eq:H_corr}
\hat{H}'/\hbar = \frac{K\varepsilon_q}{6}\Bigg{[}\Big{(}\hat{a}^2 + \hat{a}^{\dagger^2}\Big{)}\Big{(}\hat{n}-1\Big{)} + \text{h.c}\Bigg{],}
\end{equation}
up to quartic order in \autoref{eq:H_painted}.
For $|K| \varepsilon_q  \ll\omega$, this term shifts the cat size $|\alpha|$ (\autoref{eq:alpha}) to a new semiclassical value 

\begin{equation}
\pm|\alpha'| \approx \pm \sqrt{\frac{\omega \varepsilon_q/8 - \delta/2 - K\varepsilon_q/3}{|K|(1+2\varepsilon_q/3)}} \approx \pm|\alpha|\Big{(}1-\frac{\varepsilon_q}{3}-\frac{2\eta}{3}  \Big{)}, 
\end{equation}
which, for the quadratic drives ($\varepsilon_q \sim 0.2 - 0.5$) and relative anharmonicity ($\eta \sim0.01-0.05$) used here, reduces $|\alpha|$ from \autoref{eq:alpha} by $\sim7-20\%$.

In our experiments, we tune the unpainted trap depth $\tilde{V}_0$ in order to maintain a fixed value of $\omega$. With $\omega \approx \omega_{\kappa} + 2K_4 \approx \omega_{G}(1 - 3\kappa^2/2) + 2K_G(1-2\kappa^2)$ for small $\kappa$, and assuming $|K_G| \ll \omega_{G}$, this amounts to increasing ${V}_G$ as we increase $\kappa$ to maintain a $\kappa$-independent transition frequency. For convenience, given that $\omega$ is the experimentally measurable frequency,  in the main text and elsewhere we  define a reference depth $V_0 \equiv mW^2\omega^2/4$.

\subsection{Frequency-chirped quadratic drive}
\label{met:chirping}
\begin{figure*}[t]
    \centering
    \includegraphics[width=0.65\textwidth]{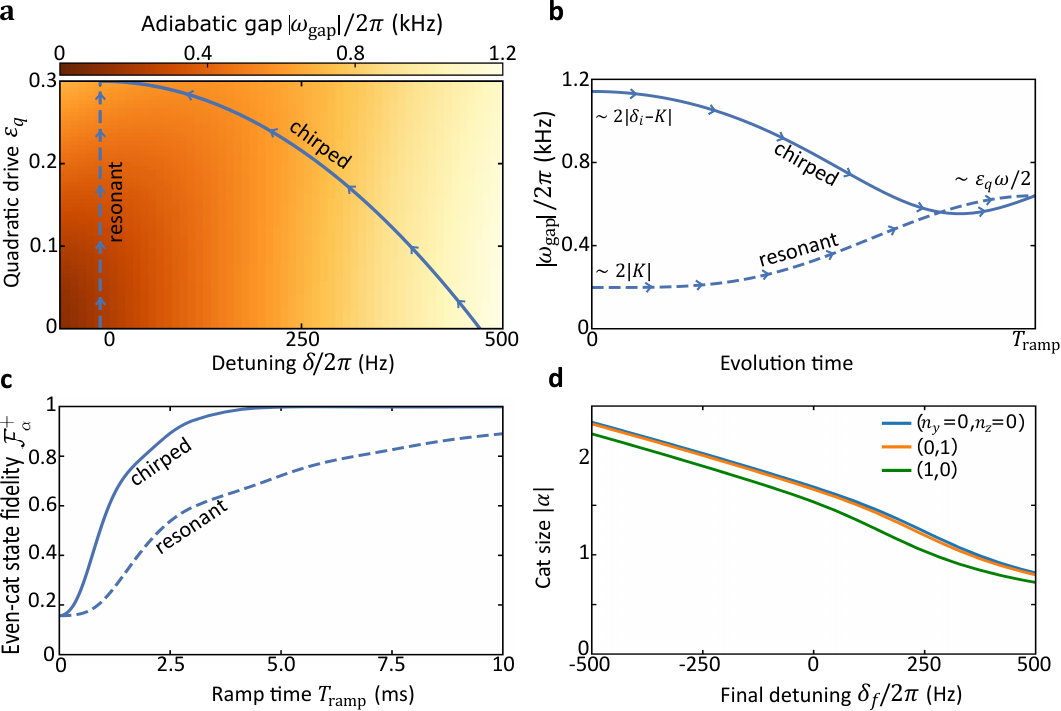}
    \caption{\textbf{Adiabatic preparation of cat states}: \textbf{a,} Adiabatic frequency gap magnitude $|\omega_{\mathrm{gap}}|$ as a function of quadratic drive amplitude $\varepsilon_q$ and detuning $\delta$. To prepare the cat states, $\varepsilon_q$ can be increased either on resonance ($\delta=0$, dashed line) or by starting from an initial blue detuning $\delta_i\equiv\delta(t=0)>0$ and chirping to resonance using a raised-cosine envelope (solid line, see Table~\ref{tab:modulation} for chirping profile). \textbf{b,} Minimum adiabatic gap between the lowest even-parity states as a function of ramp time $T_{\mathrm{ramp}}$. For $\delta=0$, the minimum gap is set by $|\omega_{\mathrm{gap}}|\sim 2|K|$. By chirping, this initial gap can be increased in proportion to $2|\delta_i-K|$. \textbf{c,} Even-parity cat-state fidelity $\mathcal{F}_{\alpha}^{+}$ (\autoref{eq:state_fidelity}) as function of $T_{\mathrm{ramp}}$ for the chirped and resonant path. $\mathcal{F}^{+}_{\alpha\approx1.5}(T_{ramp}=0)$ corresponds to the overlap $|\bra*{n = 0}\ket*{\mathcal{C}^{+}_{\alpha=1.5}}|^2 \approx 0.15$. \textbf{d,} Cat size $|\alpha|$ as a function of the final detuning $\delta_f\equiv\delta(t=T_{\mathrm{ramp}})$, after a perfect adiabatic ramp, for different motional populations in the radial and axial modes $(n_y,n_z)$ orthogonal to the cat dimension $x$. Although non-zero population shifts the nominal value of $|\alpha|$, the resulting reduction in state fidelity is negligible ($\mathcal{F}^{+}_{\alpha(n_y,n_z)}>99\%$ for the detuning values shown). In all plots we consider a painted-potential with $\omega = 2\pi \times \SI{8}{kHz}, 2K = 2\pi \times \SI{200}{Hz}$.}
    \label{fig:adiabatic_ramps}
\end{figure*}

Even-parity cat states are prepared using the adiabatic protocol described in section \textit{\nameref{sec:encoding}}. Adiabaticity requires $\varepsilon_q$ to be ramped slowly relative to the instantaneous energy gap $\omega_{\mathrm{gap}}$ between the ground and first excited even-parity eigenstates. Figure~\ref{fig:adiabatic_ramps}a shows this gap as a function of $\varepsilon_q$ and a quadratic-drive detuning $\delta$. For $\delta=0$, the gap increases monotonically from an initial value of $2K$ to $4K|\alpha|^2=\omega\varepsilon_q/2$. Since $K$ is the smallest energy scale in the system, it sets the minimum gap and therefore the maximum adiabatic ramp speed.
However, this limitation can be overcome by frequency chirping~\cite{qing2026quantum}. Beginning with an initial detuning $\delta_i\equiv\delta(t=0)$ blue of the quadratic resonance, chirping to a final detuning $\delta_f\equiv\delta(t=T_{\mathrm{ramp}})$ while simultaneously increasing $\varepsilon_q$ enlarges the initial gap between $\ket{0}$ and $\ket{2}$ to approximately $\omega_{\mathrm{gap}}\propto-2\delta_i+2K$ (\autoref{fig:adiabatic_ramps}a,b). Consequently, higher adiabatic preparation fidelities can be achieved with substantially shorter ramp durations (\autoref{fig:adiabatic_ramps}c).

\subsection{Matrix elements of the Kerr-cat and Fock Hamiltonians}\label{met:matrix_Hamiltonians}
\quad \textit{Kerr cat ---} For large $|\alpha|$, where $\alpha$ is assumed to be real, the matrix elements of the linear drive in the basis of the low-lying states for~\autoref{eq:KCHamiltonian} $\{ \ecat, \ocat, \execat, \exocat\}$ are given by
\begin{align}
    \hat{H}_{q,l}^{\mathcal{C}}/\hbar \approx \begin{pmatrix}
        0 & \omega \varepsilon_l \alpha/2 & 0 & \omega \varepsilon_l/4 \\
        \omega \varepsilon_l \alpha/2 & 0 & \omega \varepsilon_l/4 & 0\\
        0 & \omega \varepsilon_l/4 & \omega \varepsilon_q/2 & 0 \\
        \omega \varepsilon_l/4 & 0 & 0 & \omega \varepsilon_q/2
    \end{pmatrix}.
\end{align}

In order to avoid leakage outside the ground cat-state doublet, we require that $\omega \varepsilon_l/4 \ll \omega_{\mathrm{gap}} = \omega \varepsilon_q/2=4K|\alpha|^2 $. Hence for $\varepsilon_l \ll 2\varepsilon_q$ the target subspace is protected. Here, our Hamiltonian can be written as $\hat{H}/\hbar = \Omega \hat{\sigma}_{x}$ where $\hat{\sigma}_{x} = \ecat\bra{\mathcal{C}^{-}_{\alpha}} + \ocat\bra{\mathcal{C}^{+}_{\alpha}}$ and the Rabi frequency is given by  $\Omega = \omega \varepsilon_l |\alpha|/2$.
Hence, the ground cat-state parity-inversion time is $T_{X} = {\pi}/{(|\alpha| \varepsilon_l \omega) }$.

\textit{Fock---}To generate excited Fock states, we modulate the trap depth and position according to \autoref{eq:linear_drive} and \autoref{eq:quadratic_drive}. Setting $\delta=0$ in \autoref{eq:KCHamiltonian} with $\varepsilon_q=0$ and $\varepsilon_l\neq0$, the Hamiltonian in the Fock basis $\{\ket0,\ket1,\ket2\}$ becomes
\begin{equation} \label{eq:matrix_Hamiltonian_Fock}
    \hat{H}_{l}^{|n\rangle}/\hbar = \begin{pmatrix} 0 & \omega\varepsilon_l/4 & 0 \\ \omega\varepsilon_l/4 & 0 & \omega\varepsilon_l\sqrt{2}/4 \\ 0 & \omega\varepsilon_l\sqrt{2}/4 & 2K \end{pmatrix}. \\
\end{equation}
Thus, coherent coupling between $\ket0$ and $\ket1$ can be achieved while suppressing leakage into $\ket2$ if the condition $\omega\varepsilon_l\ll 4\sqrt{2}|K|$ is satisfied. To drive the transition $\ket0\leftrightarrow\ket2$, we use only intensity modulation ($\varepsilon_q\neq0$, $\varepsilon_l=0$), in which case only parity-preserving transitions are allowed. By setting $\delta =K$, the  Hamiltonian in the Fock basis $\{\ket{0},\ket{2},\ket{4}\}$, can be written as 
\begin{equation}
    \hat{H}_{q}^{|n\rangle}/\hbar = \begin{pmatrix} 0 & \omega\varepsilon_q\sqrt{2}/8 & 0 \\ \omega\varepsilon_q\sqrt{2}/8 & 0 & \omega\varepsilon_q\sqrt{3}/4 \\ 0 & \omega\varepsilon_q\sqrt{3}/4 & 8K \end{pmatrix}, \\
\end{equation}
where leakage outside the qubit subspace is suppressed when $\omega\varepsilon_q\sqrt{3} \ll 32|K|$.

\subsection{Three-dimensional motional population and state fidelity}
\label{met:3D_pop_and_fidelity}

\begin{figure*}[t]
\centering
\includegraphics[width=0.67\textwidth]{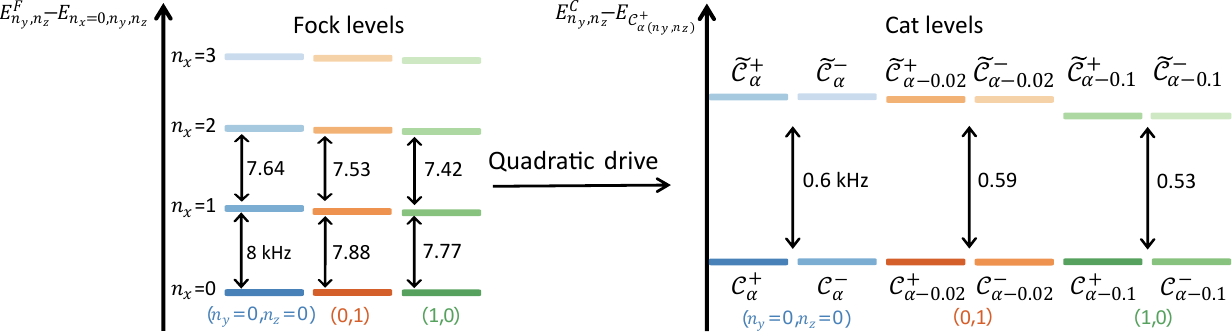}
\caption{ 
\textbf{Energy level diagrams of 3D motional states.} Energy splittings between the Fock levels and cat-state levels (all specified in kHz) in relation to motional population in orthogonal tweezer dimensions $n_y,n_z$ (indicated by different shades). When the quadratic drive is adiabatically turned on, the different Fock states are transformed into corresponding cat states with an effective cat size dependent on the occupation in $(n_y,n_z)$. While Fock state splitting are quite sensitive to this occupation, the resulting splitting between cat states is insignificant for the mode numbers used, with the shown states accounting for ~99\% of the cooled population. Here we use $\omega = 2\pi \times \SI{8}{kHz}, 2K = 2\pi \times \SI{200}{Hz}.$ }
\label{fig:level_diagram}
\end{figure*}

Due to imperfect 3D ground-state cooling, there is residual excited Fock-state population prior to the application of intensity or position modulation. This is true for both the dimension of interest ($x$) and in the orthogonal dimensions ($y,z$). Dimensional coupling terms in the 3D Hamiltonian shift the trap frequency and nonlinearity in $x$ according to the mode numbers ($n_y,n_z$): $\omega \rightarrow  \omega(n_y,n_z), K \rightarrow K(n_y,n_z)$ (see \autoref{fig:level_diagram} for 3D level diagram).
If our state is initially prepared with $(n_x,n_y,n_z)$, the trap frequency and nonlinearity will be shifted relative to the 3D ground-state as $\Delta \omega = \omega(n_y,n_z) - \omega(0,0)$ and  $\Delta K = K(n_y,n_z) - K(0,0)$, thereby renormalizing the parameters in our Hamiltonian.

For cat states, these  excitations shift the cat size as $|\alpha| \!\rightarrow \! |\alpha(n_y,n_z)|$, which for small excitations ($n_{y,z} {\sim} 1$), can be expressed as $|\alpha(n_y,n_z)| \approx |\alpha|(1+\frac{\Delta \omega}{2\omega(0,0)} - \frac{\Delta K}{2K(0,0)})$ (\autoref{fig:adiabatic_ramps}d). While each pair of states
$\{|C^+_{\alpha(n_y,n_z)}\rangle,|C^-_{\alpha(n_y,n_z)}\rangle\}$ remains degenerate, the Rabi frequency $\Omega \sim |\alpha(n_y,n_z)|$ and the gap $\omega_{\mathrm{gap}}\sim \omega(n_y,n_z)$ will be slightly reduced. Therefore, cat-state fidelity is limited primarily by initial excitations in the cat mode, and is largely insensitive to small excitations in orthogonal modes. For Fock states, non-zero population in orthogonal modes results in a sizable frequency shift (\autoref{fig:level_diagram}). Consequently, only atoms in the 3D motional ground state participate efficiently in the intended transition, placing a stricter bound on the maximum excited-state fidelity.

\subsection{Cat- and Fock-state fidelity simulations}
\label{met:trap_freq_and_fidelity}

\textit{\quad Simulations ---} In \autoref{fig:fock}f, we simulate state fidelity of the $|1\rangle$ Fock state and $|\mathcal{C}_{1.2}^{-}\rangle$ cat state in the presence of shot-to-shot frequency noise with a standard deviation of  0.35\%. Using the experimentally measured initial ground-state occupations,  state fidelity is computed using Monte Carlo simulations of the mixed-state dynamics. 
For the cat-state simulations, we neglect motional coupling between orthogonal tweezer dimensions and assume that the ramp between our initial Fock thermal-distribution and cat state is fully adiabatic. Accordingly, any population in $\ket{n_x = 0,n_y,n_z}$ is mapped to $|\mathcal{C}_{\alpha(n_y,n_z)}^{+}\rangle$,   $\ket{n_x = 1,n_y,n_z}$ is mapped to $|\mathcal{C}_{\alpha(n_y,n_z)}^{-}\rangle$, $\ket{n_x = 2,n_y,n_z}$ is mapped to $|\tilde{\mathcal{C}}^{-}_{\alpha(n_y,n_z)}\rangle$, and so on. Evolving this initial cat state  under the linear drive, we optimize all parameters such that the population in the thermal odd-cat manifold is maximized at a trap frequency $\omega_{\mathrm{opt}}$. The trap depth is then decreased, yielding a reduced trap frequency $\omega$, and the drive pulse optimized for $\omega_{\mathrm{opt}}$ is applied. The state fidelity is then calculated for different fractional trap frequencies $\omega/\omega_{\mathrm{opt}}$. Note that for the Fock state simulations, full 3D models incorporating cross-mode couplings are used.

\textit{Trap frequency noise ---}
Here, we examine the impact of shot-to-shot trap frequency fluctuations and initial ground-state population on state fidelity (\autoref{fig:fock}e). When the trap frequency changes ($\omega \rightarrow \omega + \delta$), the effective $|\alpha|$ of the cat state changes by $2|\alpha| \delta/\omega \varepsilon_q$ (\autoref{eq:alpha}) to leading order. Taking the parity inversion time  corresponding to the unshifted case, $T_{X} = \pi/(\omega \varepsilon_q |\alpha|)$, the odd cat state population is given by $P^{-}_{\alpha} = \sin^2(\frac{\pi}{2}(1+\frac{2\delta}{\omega \varepsilon_q})) \approx 1-(\frac{\pi \delta}{\omega \varepsilon_q })^2$. Thus, although trap-frequency shifts modify the Rabi frequency, shot-to-shot fluctuations of only $\delta = 0.35\%\times\omega$ produce negligible fidelity loss ($\sim0.1\text{--}0.3\%$). For Fock states, a shift in trap frequency detunes the transition by $\delta$. The population of $\ket{1}$  at the optimized parity-inversion time $T_{X} = 2\pi/\varepsilon_l \omega$ is then $P^{\ket{1}} \approx 1-(2{\delta}/{\omega \varepsilon_l})^2$. In this idealized scenario, the Fock-state sensitivity to detuning is larger than that of the cat state by a factor of $[2\varepsilon_q/(\pi\varepsilon_l)]^2$, which for the parameters used in \autoref{fig:fock}e, is a factor of $\sim$40. In both experiment and simulation, however, we find only an $\sim$8-fold reduction in detuning sensitivity. This is likely smaller than the predicted value due to imperfect preparation of $|0\rangle$ and trap-frequency noise. We also note that the semiclassical expression for $|\alpha|$, given above, overestimates the odd cat state fidelity by $\sim 3\%$ at $\omega/\omega_{opt} = 0.97$ compared to exact diagonalization of the Kerr-cat Hamiltonian (\autoref{eq:KCHamiltonian}) with correction (\autoref{eq:H_corr}).

\subsection{Experimental apparatus}\label{met:apparatus}

A simplified schematic of the experimental apparatus is shown in \autoref{fig:expdiagram}(a); a detailed description is given in Ref.~\cite{brown2022light}. Briefly, optical tweezers are generated by directing \SI{852}{\nano\meter} laser light through two orthogonal longitudinal-mode TeO$_2$ acousto-optic deflectors (AODs: IntraAction ATD-1803DA2.850 with access time \SI{235}{\nano\second\per\milli\meter}), driven by an arbitrary waveform generator (NI PXIe-5451). While the tweezer power is regulated by an intensity control loop for trap depths between approximately \SI{500}{\micro\kelvin} and \SI{2}{\milli\kelvin}, shallower traps are realized by controlling the relative power in the radio-frequency (RF) tones driving the AOD. With this approach, up to 99\% of the remaining optical power is transferred to an auxiliary tweezer positioned $\sim$\SI{8}{\micro\meter} from the main tweezer. For deep traps ($\sim\si{\milli\kelvin}$), the  trap depth $V_0$ is determined from the beam-induced light shift, while $\omega$ is measured independently using Raman spectroscopy or parametric heating~\cite{kaufman2012cooling}. These independent measurements can be used to estimate the  beam waist $W \approx \sqrt{4V_0/(m\omega^2)}$. For shallow traps ($\sim$\SI{}{\micro\kelvin}), $\omega$ and $K$ are determined from displaced-state measurements. For readout, atomic fluorescence is collected on an EMCCD camera (Andor iXon 897 Ultra).

\subsection{Experimental timing}\label{met:timing}

A schematic of the full experimental timing sequence is shown in Fig. \ref{fig:expdiagram}b. We begin by loading a single $^{87}$Rb atom from a magneto-optical trap (MOT) into an optical tweezer at a depth of $V_0/k_B\approx\SI{2}{\milli\kelvin}$ with a loading probability of $\sim60 \%$. Atom presence is verified for post-selection by collecting fluorescence during a \SI{15}{\milli\second} in-trap imaging period, taken at a reduced depth of $V_0/k_B\approx\SI{500}{\micro\kelvin}$. Loading and imaging are performed using $\sigma^+ \sigma^-$ polarization gradient cooling (PGC) with light red-detuned from the $F=2 \rightarrow F'=3$ transition.

After imaging and additional PGC,  Raman sideband-cooling~\cite{kaufman2012cooling} is performed sequentially along the radial and axial directions of a trap at a depth of $V_0/k_B\approx\SI{1}{\milli\kelvin}$ with radial (axial) trap frequency   \SI{135}{\kilo\hertz} (\SI{40}{\kilo\hertz}). For each direction, 150 Gaussian-shaped pulses are applied, resulting in a total Raman cooling duration of $\sim$\SI{33}{\milli\second}. After cooling, we measure the 3D ground state using Raman sideband spectroscopy, yielding an initial occupation of $\{ \bar{n}_x, \bar{n}_y, \bar{n}_z \} = \{ 0.035(35), 0.035(35), 0.087(60)\}$, corresponding to a 3D ground state fraction of $86(6)\%$. These values represent an average from several measurements throughout the experimental campaign; the quoted uncertainty intervals are the standard deviations of the measurement distributions. Following Raman cooling, the trap depth is adiabatically reduced to $V_0/k_B\approx\SI{4}{\micro\kelvin}$ for motional state preparation as described in the text. Finally, the quantum state is measured using time-of-flight (TOF) tomography.

\subsection{Trap modulation}
\label{met:mod_methods}
To realize painted potentials, the RF tone driving the AOD that controls the tweezer position is sinusoidally modulated. A modulation frequency of $\SI{1}{MHz} \gg \omega/2\pi$  ensures a time-averaged potential free of parametric heating while also remaining within the finite operational bandwidth of the AOD. To prevent motional excitation during initialization, the modulation amplitude is smoothly ramped from zero to its steady-state value over a duration of $\SI{1}{ms}$ using a hyperbolic tangent envelope. The linear drive is implemented using the same technique, but at a reduced frequency and amplitude. We note that when applying the linear drive, we observe a small initial displacement that depends on the center frequency of the RF tone driving the AOD. The origin of this noise remains unknown and possibly limits the fidelity of larger odd-parity cat states, which are highly sensitive to such displacements (see \autoref{fig:cats}d).

The quadratic drive is implemented by sinusoidally transferring optical power between the main and auxiliary tweezers through modulation of the relative RF tone amplitudes driving the AOD. The quadratic-drive frequency is chirped using a raised-cosine envelope, beginning with an initial detuning of $\delta_i\approx2\pi\times\SI{1}{kHz}$ (blue-detuned from the $\omega_q/2\pi$ resonance). When ramping from cat state to Fock state (\autoref{fig:fock}f), this ramp is inverted. Although more elaborate frequency trajectories could be implemented using alternative envelope profiles, we found the raised-cosine chirp to be sufficient for $\delta_i/2\pi\approx0.3\text{–}2~\mathrm{kHz}$ and ramp durations of $T_q\sim3\text{–}10~\mathrm{ms}$.

\subsection{Displaced ground-state measurements}\label{met:coherent_state}

To probe the energy landscape of the motional states (\autoref{fig:kerr_figure}), we use a spectroscopy protocol based on TOF imaging. For the Hamiltonian $\hat{H}_p$ corresponding to the painted potential (\autoref{eq:H_painted}), let $\ket{\psi_n}$ and $E_n$ denote its eigenstates and eigenenergies. Starting from a displaced ground state represented as $\ket{\Psi}=\sum_n c_n \ket{\psi_n}$, we allow the system to evolve for a time $t_e$ and measure the mean momentum $\langle \hat{p}(t) \rangle$ via TOF. The measured signal in the eigenbasis is
\begin{align}
\begin{split}
    \langle \hat{p}(t) \rangle &= \bra{\Psi} e^{i\hat{H}_pt/\hbar} \hat{p} e^{-i\hat{H}_pt/\hbar}\ket{\Psi}, \\
    &= \sum_{n,m} c_nc_m^* \bra{\psi_m}\hat{p}\ket{\psi_n} e^{-i(E_n-E_m)t}. \\
\end{split}
\end{align}
Here $\hat{p}(t)$ represents the momentum operator in the Heisenberg picture. Since the painted potential remains close to a harmonic potential, the momentum matrix elements are approximately $\bra{\psi_m}\hat{p}\ket{\psi_n}\simeq ip_0\sqrt{n+1} \delta_{n,m-1}-ip_0\sqrt{n}\delta_{n,m+1}$, and
\begin{equation}
    \langle \hat{p}(t) \rangle \simeq \sum_n ip_0\sqrt{n+1}c_nc_{n+1}^* e^{i(E_{n+1}-E_n)t} + c.c., \\
\end{equation}
showing that the Fourier spectrum of the TOF signal directly reveals the transition frequencies $E_{n+1}-E_n$ between neighboring eigenstates.

After ground-state cooling, the displacement is realized by abruptly shifting ($t_{\mathrm{shift}} \ll 2\pi/\omega$) the trap by $\sim x_0$. After the atom then undergoes phase-space evolution within the anharmonic potential, the trap is rapidly extinguished and $\langle \hat{p}(t) \rangle$ is extracted via fluorescence imaging following free-expansion for a flight-time $t_f=150-250 \ \mu$s. The resulting time-domain measurements are fit to a model consisting of two sinusoids with an exponentially decaying envelope,
\begin{equation}
y(x) = e^{-\gamma x} \left[ A_1 \sin(\omega_{01} x + \phi_1) + A_2 \sin(\omega_{12} x + \phi_2) \right], 
\end{equation}
where $\gamma$ is the common decay rate, $ \omega_{01}=\omega$ and $\omega_{12}$ are the transition frequencies of the two lowest motional levels, $A_1$ and $A_2$ are the respective frequency amplitudes, and $\phi_1$ and $\phi_2$ are the respective phases. Parameter uncertainties are $1\sigma$ error of the fit. 

By simulating the evolution of the displaced thermal state with average initial occupations ${\bar{n}_x,\bar{n}_y,\bar{n}_z}$, we determine the experimental parameters that best fit the data through $\chi^2$ minimization. While our measurements cannot discern the origin of the decaying envelope seen in \autoref{fig:kerr_figure}, it can be explained by assuming static shot-to-shot tweezer intensity fluctuations with a fractional amplitude of 0.7\%, corresponding to a 0.35\% fractional change in $\omega$.

To obtain the theoretical curves in \autoref{fig:kerr_figure}c, we calculate the motional energy levels of an $^{87}\text{Rb}$ atom confined within a painted potential given by \autoref{eq:time_averaged_potential}. The motional eigenvalues are obtained by numerically solving the time-independent Schrödinger equation, and provide the transition frequencies and the absolute anharmonicity $2K$ as a function of the modulation amplitude $\kappa$.

\subsection{Tomography analysis}\label{met:tomography}

The TOF tomography technique employed in this work follows the method used in Ref.~\cite{brown2023time}, which we briefly summarize here. Analogous to optical homodyne tomography, the state is reconstructed from quadrature probability distributions measured at various phase-space angles. After target state preparation, the atom evolves in the optical tweezer for a variable time $t_e$, corresponding to a phase-space rotation by $\theta=t_e\omega_{\mathrm{eff}}$, where $\omega_{\mathrm{eff}}\approx\omega+2K\bar{n}$ is the effective trap frequency. After the state has rotated by the desired $\theta$, the tweezer is rapidly extinguished and the atom undergoes free-space expansion for a duration $t_f$. This maps its momentum distribution at $\theta$ onto a spatial distribution, which is measured via polarization-gradient fluorescence imaging in the y–z plane (\autoref{fig:expdiagram}a) for a duration of \SI{15}{\micro\second}. Repeating this sequence approximately $10^4$ times per $\theta \in [0, \pi]$ yields the quadrature probability distribution $P(\tilde{p}, \theta)=\mathrm{tr}\left( \hat{\rho} \left| \tilde{p}, \theta \right\rangle \left\langle\tilde{p}, \theta \right|\right)$ where $\left| \tilde{p}, \theta \right\rangle$ are the momentum-space harmonic oscillator wavefunctions rotated by $\theta$. The number of measured quadrature angles is chosen to exceed the effective Fock-space dimension of the reconstructed state, ensuring informational completeness of the tomography in the truncated Hilbert space~\cite{sych2012informational}.

For each $\theta$, images containing a loaded atom are averaged and the mean of the unloaded background images is subtracted. The resulting average image is deconvolved with the measured point-spread function (PSF) using a Lucy–Richardson algorithm filtered by a thresholding parameter $\zeta$, chosen by a self consistency bootstrap \cite{brown2023time}. Integrating the deconvolved image along the $y$-axis yields the 1D quadrature distribution along $x$, $P_x(\tilde{p}/p_0,\theta)\equiv P(\tilde{p}/p_0,\theta)$. Pixel-to-momentum mapping is determined from the pixel position relative to the trap location (calibrated from in-trap atom images), scaled by the imaging magnification and expansion time, and corrected for gravitational free fall over that same expansion time. The resulting quadrature distribution and $\omega_{\mathrm{eff}}$ are used in a maximum-likelihood estimation (MLE) algorithm to reconstruct the density matrix $\hat{\rho}_{\mathrm{rec}}$.

\subsection{Image analysis parameters}\label{met:image_analysis}

The camera settings and imaging parameters used throughout this work are summarized in \hyperref[tab:imaging]{Table 1}. Both the in-trap and TOF PSFs are obtained by fitting a two-dimensional Gaussian function to an averaged image of single-atom measurements. For TOF, the PSF is measured immediately following the release of a ground-state atom from the trap. Two TOF imaging configurations with different PSFs were used in this work. In one configuration, a quantization axis was maintained in the $x$-direction during imaging; this configuration was used for the Fock-state data in \autoref{fig:fock}a-d and the cat-state data in \autoref{fig:cats}b,d. For all remaining states, we zero the magnetic field prior to the TOF interval, giving rise to a slightly smaller imaging PSF in $y$. For both TOF PSFs, an $8\%$ correction is included to account for the atom's zero-point motion. Further reduction in the PSF may be possible by applying a sequence of short imaging pulses with alternating propagation directions, thereby minimizing radiation pressure along any one direction~\cite{bergschneider2019experimental}. While often necessary for atoms with small mass, such an approach was not used here due to the relatively large mass of ${}^{87}\mathrm{Rb}$. 

The magnification of the imaging system was calibrated using an atom-drop measurement, following the procedure described in the Supplementary Material of Ref.~\cite{brown2023time}. Here, atoms are released from the trap and allowed to fall under gravity for varying durations before imaging. The resulting displacement in camera pixels is then compared with the known gravitational displacement. Magnification and PSF uncertainties correspond to $1\sigma$ error of the fit.

\subsection{Statistical errors through bootstrapping}\label{met:bootstrap}

For each reconstructed quantum state, we quantify the reconstruction accuracy using the fidelity,
\begin{equation}
\label{eq:state_fidelity}
\mathcal{F}(\hat \rho_\mathrm{t}, \hat \rho_\mathrm{rec}) =
\left(\mathrm{tr}\sqrt{\sqrt{\hat \rho_\mathrm{t}}
\hat \rho_\mathrm{rec}
\sqrt{\hat \rho_\mathrm{t}}}\right)^2,
\end{equation}
where $\hat \rho_\mathrm{t}$ is the target state and $\hat \rho_\mathrm{rec}$ is the reconstructed state. The target state depends on the state being reconstructed. For cat states, $\hat \rho_\mathrm{t}$ is taken to be an ideal cat state with coherent-state amplitude $|\alpha|$ determined via $\langle \hat{a}^2\rangle=\mathrm{tr}(\hat{a}^2\hat \rho_\mathrm{rec})=|\alpha|^2$. For Fock states, $\hat \rho_\mathrm{t}$ is the corresponding ideal Fock state. As we did not target a specific rotation angle, $\hat \rho_\mathrm{t}$ for the Fock-state superpositions is taken to be an ideal pure state whose populations and relative phase are extracted from $\hat \rho_\mathrm{rec}$.

To quantify statistical uncertainties in the reconstructed quantum states, we use a non-parametric bootstrap method. Here, $\hat \rho_{\mathrm{rec}}$ is first reconstructed from $P(\tilde{p}/p_0,\theta)$ via maximum-likelihood estimation (MLE). For each $\theta$, a bootstrapped data set is generated by randomly resampling the measured images with replacement to produce a new average quadrature distribution. Each resampled data set is then processed through the same tomography pipeline used for the experimental data. Repeating this procedure generates an ensemble of $N$ bootstrapped density matrices, $\{\hat \rho_{\mathrm{boot}}\}$, from which statistical uncertainties are estimated. To estimate the statistical uncertainty with approximately $0.1\%$ precision, we use $N=200$ bootstrap samples.

\begin{figure*}[p] 
\centering

  \includegraphics[width=1\textwidth]{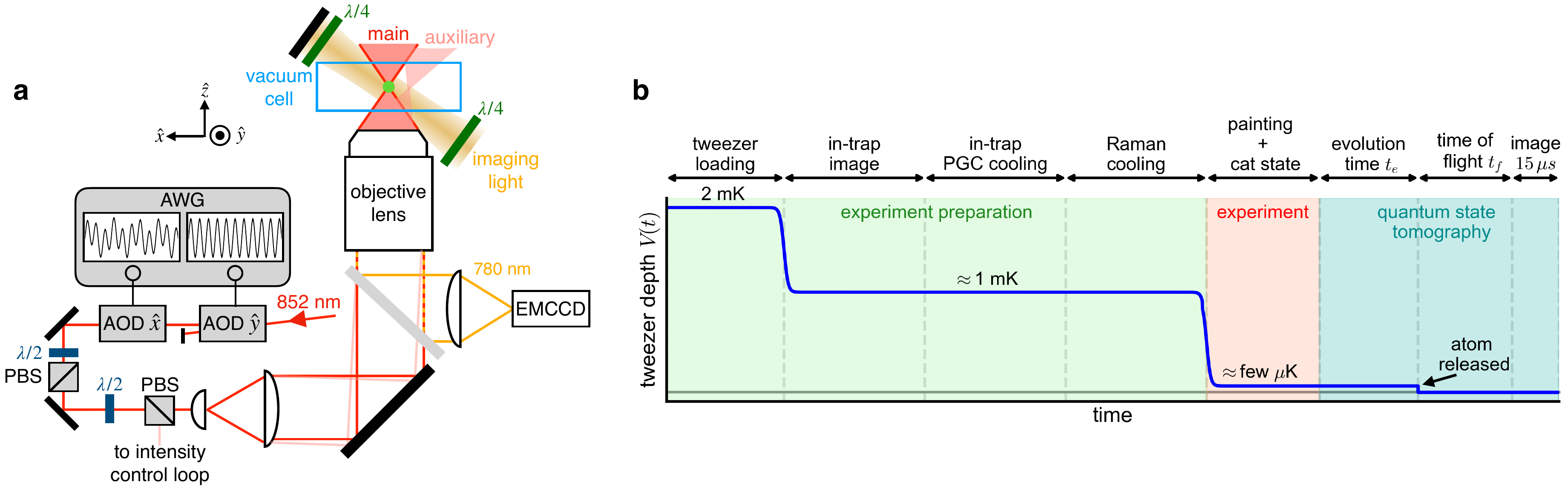}
  \caption{ 
  \textbf{Schematic experimental and timing diagram} \textbf{a,} Simplified experimental diagram. An arbitrary waveform generator (AWG) drives a pair of orthogonal acousto-optic deflectors (AODs) that control position and relative power of the main and auxiliary tweezers. After passing through a series of half-waveplates ($\lambda/2$) and polarizing beam splitters (PBS), the light is focused through an objective lens with NA$\approx$ 0.5 to a waist of $W\approx \SI{0.7}{\micro m}$. Atom fluorescence, excited by ${\sim}\SI{780}{nm}$ light in a $\sigma^+\sigma^-$ polarization gradient configuration (imaging light), is collected through the same objective and reflected off a dichroic mirror onto an electron-multiplying charge-coupled device (EMCCD) camera. In our system, the gravitational acceleration is directed along the $\hat{x}$-axis. \textbf{b,} Not-to-scale timing sequence of the optical tweezer experiment. The blue trace shows the approximate trap depth during a single experimental repetition. The shaded regions denote experiment preparation (green), the experimental sequence—including tweezer painting, state preparation, and coherent control (red)—and quantum state tomography (blue). Unless otherwise indicated, all trap-depth changes are performed using \SI{1}{\milli\second} hyperbolic-tangent adiabatic ramps.
  }
  \label{fig:expdiagram}

  \vspace{2.5em} 

  \makebox[\textwidth][c]{%
    \begin{minipage}[t]{0.45\textwidth}
      \centering
      
    {\fontsize{8.5}{9.5}\selectfont
    \noindent\textbf{Table 1} Time-of-flight imaging parameters.\\
    }
      \label{tab:imaging}
      \vspace{5pt}
      
      \begin{tabular}{ll}
      \hline\hline
      \textbf{Parameter} & \textbf{Value} \\
      \hline
      \multicolumn{2}{l}{\textit{Camera settings}} \\[3pt]
      Vertical shift speed & \SI{2}{\mega\hertz} \\ 
      Horizontal shift speed & \SI{1}{\mega\hertz} \\
      EM gain & $\times 350$ \\
      Pixel size & \SI{16}{\micro\meter} \\
      Pixel binning & $1 \times 1$ \\
      Camera temperature & \SI{-60}{\celsius} \\[5pt]
      \multicolumn{2}{l}{\textit{Imaging parameters}} \\[3pt]
      TOF imaging time & \SI{15}{\micro\second} \\
      TOF PSF ($B\neq0$) $(\sigma_x,\sigma_y)$ & $[1.47(5),\,2.40(7)]$ pixels \\
      TOF PSF ($B=0$)& $[1.53(7),\,2.08(8)]$ pixels\\
      In-trap imaging time & \SI{15}{\milli\second} \\
      In-trap PSF $(\sigma_x,\sigma_y)$ & $[1.19(6),\,1.15(6)]$ pixels \\
      Magnification & $67(3)$ \\
      Camera conversion factor $C_{\mathrm{ph}}$ & \SI{94.3}{counts/photon} \\
      Collected photons per atom & $\sim10$ \\
      \hline\hline
      \end{tabular}
    \end{minipage}
    \hfill 
    
    \begin{minipage}[t]{0.50\textwidth}
      \centering
      \setlength{\tabcolsep}{6pt}   

    {\fontsize{8.5}{9.5}\selectfont
    \noindent\textbf{Table 2} States and fidelities $\mathcal{F}(\hat{\rho}_{t},\hat{\rho}_{rec})$ with uncertainties corresponding to the central 68.2\% bootstrap confidence interval. All values are percentages. Here, $\mathrm{tr}\left(\hat{\rho}_{\mathrm{rec}}^{\mathrm{sub}}\right)$ is the trace over the relevant subspace states.\par
    }
      \label{tab:fidelity_summary}
      
      \renewcommand{\arraystretch}{1.2} 
      \vspace{5pt}
      
      \begin{tabular}{lcc}
      \hline\hline
      \textbf{State} &
    $\pmb{\mathcal{F}}(\pmb{\hat{\rho}}_\mathbf{t}, \pmb{\hat{\rho}}_\mathbf{rec})$ &
    $\pmb{\mathrm{tr}}\left( \pmb{\rho}_\mathbf{rec}^\mathbf{sub} \right)$\\
      \hline
      $\ket{\mathcal{C}_{1.8}^{+}}$ & $94.4^{+2.9}_{-2.7}$ &  $97.3^{+0.4}_{-0.4}$\\
      $\ket{\mathcal{C}_{1.8}^{-}}$ & $74.8^{+1.9}_{-1.9}$ &  $93.1^{+1.4}_{-1.3}$\\
      $\ket{\mathcal{C}_{0.9}^{+}}$ & $95.4^{+1.8}_{-1.5}$ &  $99.0^{+1.0}_{-1.7}$\\
      $\ket{\mathcal{C}_{1.3}^{+}}$ & $80.5^{+1.6}_{-1.5}$ &  $95.7^{+1.0}_{-1.0}$\\
      $\ket{\mathcal{C}_{2.1}^{+}}$& $70.8^{+1.6}_{-1.5}$ &  $97.7^{+0.7}_{-0.7}$\\
      $\ket{\mathcal{C}_{2.7}^{+}}$ & $57.3^{+1.9}_{-1.9}$ &  $86.8^{+1.9}_{-2.0}$\\
      $\ket{\mathcal{C}_{1.2}^{-}}^{\Box}$ & $90.4^{+2.1}_{-1.9}$ &  $97.6^{+0.5}_{-0.4}$\\
      $\ket{\mathcal{C}_{1.1}^{-}}^{\Diamond}$ & $80.9^{+1.9}_{-1.7}$ & $97.1^{+0.7}_{-0.7}$\\
      $\ket{\mathcal{C}_{1.0}^{-}}^{\triangle}$ & $70.1^{+2.9}_{-3.5}$ & $96.4^{+1.3}_{-0.6}$\\
      $\ket{1}^{\Box}$ & $78.0^{+2.7}_{-3.0}$ &  $95.0^{+0.9}_{-0.9}$\\
      $\ket{1}^{\Diamond}$ & $16.3^{+1.1}_{-1.1}$ &  $98.6^{+0.3}_{-0.3}$\\
      $\ket{1}^{\triangle}$ & $5.7^{+1.0}_{-0.9}$ &  $99.7^{+0.3}_{-0.2}$\\
      $\ket{1}^{\heartsuit}$ & $95.4^{+1.2}_{-1.4}$ & $96.4^{+1.0}_{-1.1}$\\
      $\ket{1}$ & $75.2^{+1.4}_{-1.5}$ & $98.8^{+0.3}_{-0.3}$\\
      $\propto\ket{0}+\ket{1}$ & $83.5^{+1.6}_{-1.6}$ &  $99.6^{+0.1}_{-0.1}$\\
      $\ket{2}$ & $57.9^{+1.0}_{-1.0}$ & $85.5^{+1.0}_{-0.9}$\\
      $\propto\ket{0}+\ket{2}$ & $75.0^{+1.3}_{-1.3}$ & $88.2^{+1.3}_{-1.3}$\\
      \hline\hline
      \end{tabular}
      \vspace{4pt}
      
      \footnotesize
      \begin{tabular}{@{}l@{}}
      $^{\Box}$ Resonantly prepared state.\\
      $^{\Diamond}$ Prepared at a depth corresponding to a 100 Hz detuning.\\
      $^{\triangle}$ Prepared at a depth corresponding to a 200 Hz detuning.\\
      $^{\heartsuit}$ $\ket{0} \rightarrow \ket{C^+_\alpha} \rightarrow \ket{C^-_\alpha} \rightarrow \ket{1}$ protocol.
      \end{tabular}
    \end{minipage}
  }
\end{figure*}

The measured fidelities are summarized in \hyperref[tab:fidelity_summary]{Table 2}. To avoid potential bias introduced by the bootstrap method, the reported fidelity is calculated directly from the experimentally reconstructed state $\mathcal{F}(\hat \rho_\mathrm{t},\hat \rho_\mathrm{rec})$. Statistical uncertainties are obtained from the fidelity distribution over the bootstrap ensemble and are reported as the central $68.2\%$ confidence interval.

\subsection{Tomography measurement error}\label{met:TOF_sim}

The estimated cat size $|\alpha|$ has a systematic uncertainty arising from the calibration of the imaging magnification $M$ and trap frequency $\omega$. Because $|\alpha| \propto 1/(M\sqrt{\omega})$, standard error propagation gives

\begin{equation}
  \frac{\sigma_{|\alpha|}}{|\alpha|} = \left( \left( \frac{\sigma_M}{M} \right)^2 + \left( \frac{\sigma_\omega}{2\omega} \right)^2\right)^\frac{1}{2},
\end{equation} 
where $\sigma_i$ denotes the uncertainty in parameter i. Using $M=67\pm3$  and $\omega/2\pi=\SI{8.0\pm0.01}{\kilo\hertz}$, we obtain $\sigma^{\mathrm{sys.}}_{|\alpha|} /|\alpha|\approx 4\%$. The statistical uncertainty in $|\alpha|$, estimated from the bootstrapped ensemble of density matrices, is approximately $\sigma^{\mathrm{stat.}}_{|\alpha|} /|\alpha|\approx3\%$. Adding these two uncertainties in quadrature we obtain a total uncertainty of $\sigma^{\mathrm{tot.}}_{|\alpha|} /|\alpha|\approx5\%$.

The accuracy of TOF tomography depends on the quantum state, as higher-energy states exhibit finer phase-space features that require greater resolution in the measured quadrature distribution $P(\tilde{p}/p_0,\theta)$. This is particularly important for cat states, whose parity is encoded in the interference fringes: an ideal cat state exhibits fringes with zero-valued minima, whereas admixture of the opposite-parity cat state reduces the fringe contrast. Finite imaging resolution arising from the finite expansion time, imaging point-spread function (PSF), system magnification, and additional sources of image blurring, can therefore cause even an ideal cat state to appear partially mixed by washing out these fringes. 

Because the PSF and magnification are fixed, we optimize the expansion time to balance momentum resolution against image blurring. For the states $|\mathcal{C}_{1.8}^{+}\rangle$ and $|\mathcal{C}_{2.1}^{+} \rangle$, we used $t_f=\SI{0.75}{\milli\second}$ for the two quadratures nearest the interference fringes ($\theta=0,\pi$), and $t_f=\SI{0.5}{\milli\second}$ elsewhere. For the state $|\mathcal{C}_{1.8}^{-} \rangle$, only the  $\theta=0,\pi$ were measured at \SI{0.75}{\milli \second} (others measured at $t_f=\SI{0.5}{\milli\second}$), possibly contributing to a reduction in state fidelity relative to $|\mathcal{C}_{1.8}^{+}\rangle$. For the state $|\mathcal{C}_{2.7}^{+}\rangle$, we use $t_f=\SI{1}{\milli\second}$ at $\theta=0,\pi$ and $t_f=0.4-\SI{0.5}{\milli\second}$ elsewhere. The states $|\mathcal{C}_{0.9}^{+}\rangle$ and $|\mathcal{C}_{1.3}^{+}\rangle$ in \autoref{fig:cats}e and the Fock states in \autoref{fig:fock}(a--d) were measured with $t_f=\SI{0.5}{\milli\second}$, whereas the Fock states in \autoref{fig:fock}(e,f) were measured with $t_f=\SI{0.75}{\milli\second}$.

To estimate the measurement error, we use TOF tomography simulations. We generate synthetic data sets by calculating the quadrature probability distributions $P_{x,y}(\tilde{p}/p_0,\theta)$ from the input density matrices $\hat \rho_{x,y}$. For each $\theta$, projective measurements are simulated by sampling momenta, converting them to pixel locations on a camera after TOF, and drawing detected photon numbers $n_{\mathrm{ph}}$ from a Poisson distribution with mean 10. Photons are distributed among pixels according to the PSF and converted to counts using the calibrated conversion factor $C_{\mathrm{ph}}$. Each pixel is then assigned a background count sampled from the same camera column of a randomly selected image from an ensemble of approximately $1.3\times10^5$ experimentally acquired background images. Repeating this procedure for $10^4$ measurements at each quadrature angle yields a synthetic tomography data set.

Using the TOF tomography simulations, we quantify the apparent fidelity loss due to finite imaging resolution as a function of $|\alpha|$. As $|\alpha|$ increases, the interference fringes become more closely spaced and progressively harder to resolve, reducing the reconstruction fidelity. The simulations assume a pure cat state along the measured dimension and the same imaging parameters as the experiment. For both even and odd cat states, the average measurement fidelity exceeds $95\%$ for $\alpha \leq 1.5$, decreases below $90\%$ at $\alpha\approx2.2$, and reaches approximately $\approx75\%$ at $\alpha=3$.

\subsection{Pulse parameters}\label{met:pulse_params}

Table 3 summarizes the pulse parameters, trap parameters, and corresponding drive equations for each state reported in this manuscript.

\section{Acknowledgments}

We thank Nicholas Frattini, Piotr Grochowski, Oriol Romero-Isart, Aashish Clerk, Matteo Marinelli, Zhenpu Zhang, Brendan Marsh, Rodrigo Cortinas, Andras Gyenis, Othmane Benhayoune-Khadraoui, and Luca Talamo for valuable input and discussion. This work was supported by the NSF JILA-PFC PHY-2317149, NSF QLCI award OMA-2016244, the U.S. Department of Energy, Office of Science, National Quantum Information Science Research Centers, Quantum Systems Accelerator (Award No. DE-SCL0000121), Office of Naval Research (ONR) grant N00014-21-1-2594, NIST, and the Baur-SPIE Chair at JILA. S.~R.~M. is supported by the NSF QLCI award OMA-2120757.

\section{Data and Code Availability}
The data and code supporting this study’s findings are available
from the corresponding author upon reasonable request.

\section{Competing interests}
The authors declare no competing interests.

\section{Author contributions}
\noindent${}^{\dagger}$these authors contributed equally to this work.

\noindent*corresponding author: regal@colorado.edu

\clearpage 

\clearpage


\begin{table*}[htbp]
    \setlength{\tabcolsep}{3pt}   
    \centering
    \setcounter{table}{2}
    \refstepcounter{table}
    \label{tab:modulation}
    \small
    {\fontsize{8.5}{9.5}\selectfont
    \noindent\textbf{Table 3} Modulation parameters for each state. Position and intensity modulations $\left( x_m(t), V_m(t) \right)$ are applied with the indicated envelope functions, where $\omega_c$ is the chirping frequency. \\\par
    }
    \vspace{5pt}
    \begin{tabular}{
        >{\centering\arraybackslash}p{0.12\textwidth}
        >{\centering\arraybackslash}p{0.15\textwidth}
        >{\centering\arraybackslash}p{0.54\textwidth}
        >{\raggedright\arraybackslash}p{0.15\textwidth}
    }
    \toprule
    \textbf{State} & \textbf{\shortstack{Trap parameters \\ $\omega/2\pi$ \\ $2K/2\pi$ } }& \textbf{Modulation function} & \textbf{Modulation parameters}\\
    \midrule
    $\left|\mathcal{C}_{1.8}^{+} \right \rangle$ & \shortstack[l]{$\SI{8479(7)}{\hertz}$ \\ $\SI{-240(18)}{\hertz}$} & $V_m(t) = \mathsf{C}(t, \tau) \cdot \varepsilon_q  V_0 \cos\left(\omega_{q} t + \frac{\omega_{c} \tau}{\pi^2}\left( \sin\left(\frac{\pi t}{2\tau}\right)-1\right)\right)\cdot \mathbf{1}_{\left[0, \tau \right]}$
        & \shortstack[l]{$\omega_{q}/2\pi=\SI{16.6}{\kilo\hertz}$ \\  $\varepsilon_q=18~\%$ \\ $\tau=\SI{5}{\milli\second}$ \\  $\omega_{c}/2\pi=\SI{942.478}{\hertz}$}\\
    \midrule

    $\left|\mathcal{C}_{1.8}^{-} \right \rangle$& \shortstack[l]{$\SI{8479(7)}{\hertz}$ \\ $\SI{-240(18)}{\hertz}$}
        & \shortstack[m]{$V_m(t) = \mathsf{C}(t, \tau) \cdot \varepsilon_q V_0\cos\left(\omega_{q} t + \frac{\omega_{c} \tau}{\pi^2}\left( \sin\left(\frac{\pi t}{2\tau}\right)-1\right)\right)\cdot \mathbf{1}_{\left[0, \tau \right]}$ \\
        $+~\varepsilon_q V_0\cos\left(\omega_{q} t\right)\cdot \mathbf{1}_{\left[\tau, \tau' \right]}$ \\
        $x_m(t)=\varepsilon_lx_{0} \cos\left(\frac{\omega_{q}}{2}t\right) \cdot \mathbf{1}_{\left[\tau, \tau'\right]}$}
        & \shortstack[l]{$\omega_{q}/2\pi=\SI{16.6}{\kilo\hertz}$ \\ $\varepsilon_q=18~\%$ \\ $\tau=\SI{5}{\milli\second}$ \\ $\omega_{c}/2\pi=\SI{942.478}{\hertz}$ \\ $\varepsilon_l=4~\%$ \\
        $\tau' = \SI{722.9}{\micro\second}$ } \\
    \midrule
        
    $\left|\mathcal{C}_{0.9}^{+} \right \rangle$ & \shortstack[l]{$\SI{8192(11)}{\hertz}$ \\ $\SI{-450(35)}{\hertz}$}
        & $V_m(t) = \mathsf{C}(t, \tau) \cdot \varepsilon_q V_0\cos\left(\omega_{q} t + \frac{\omega_{c} \tau}{\pi^2}\left( \sin\left(\frac{\pi t}{2\tau}\right)-1\right)\right)\cdot \mathbf{1}_{\left[0, \tau \right]}$
        & \shortstack[l]{$\omega_{q}/2\pi=\SI{16.6}{\kilo\hertz}$ \\  $\varepsilon_q=18~\%$ \\ $\tau=\SI{5}{\milli\second}$ \\  $\omega_{c}/2\pi=\SI{942.478}{\hertz}$}\\
    \midrule
    $\left|\mathcal{C}_{1.3}^{+} \right \rangle$ & \shortstack[l]{$\SI{8192(11)}{\hertz}$ \\ $\SI{-450(35)}{\hertz}$}
        & $V_m(t) = \mathsf{C}(t, \tau) \cdot \varepsilon_q V_0\cos\left(\omega_{q} t + \frac{\omega_{c} \tau}{\pi^2}\left( \sin\left(\frac{\pi t}{2\tau}\right)-1\right)\right)\cdot \mathbf{1}_{\left[0, \tau \right]}$
        & \shortstack[l]{$\omega_{q}/2\pi=\SI{16.6}{\kilo\hertz}$ \\  $\varepsilon_q=48~\%$ \\ $\tau=\SI{5}{\milli\second}$ \\  $\omega_{c}/2\pi=\SI{942.478}{\hertz}$}\\
    \midrule
    $\left|\mathcal{C}_{2.1}^{+} \right \rangle$ & \shortstack[l]{$\SI{7379(1)}{\hertz}$ \\ $\SI{-131(10)}{\hertz}$}
        & $V_m(t) = \mathsf{C}(t, \tau) \cdot \varepsilon_q V_0\cos\left(\omega_{q} t + \frac{\omega_{c} \tau}{\pi^2}\left( \sin\left(\frac{\pi t}{2\tau}\right)-1\right)\right)\cdot \mathbf{1}_{\left[0, \tau \right]}$
        & \shortstack[l]{$\omega_{q}/2\pi=\SI{15.0}{\kilo\hertz}$ \\  $\varepsilon_q=25~\%$ \\ $\tau=\SI{5}{\milli\second}$ \\  $\omega_{c}/2\pi=\SI{942.478}{\hertz}$}\\
    \midrule
    $\left|\mathcal{C}_{2.7}^{+} \right \rangle$& \shortstack[l]{$\SI{8257(1)}{\hertz}$ \\ $\SI{-100(2)}{\hertz}$}
        & $V_m(t) = \mathsf{C}(t, \tau) \cdot \varepsilon_q V_0\cos\left(\omega_{q} t + \frac{\omega_{c} \tau}{\pi^2}\left( \sin\left(\frac{\pi t}{2\tau}\right)-1\right)\right)\cdot \mathbf{1}_{\left[0, \tau \right]}$
        & \shortstack[l]{$\omega_{q}/2\pi=\SI{15.0}{\kilo\hertz}$ \\  $\varepsilon_q=50~\%$ \\ $\tau=\SI{5}{\milli\second}$ \\  $\omega_{c}/2\pi=\SI{942.478}{\hertz}$}\\
    \midrule
    $\left|\mathcal{C}_{1.2}^{+} \right \rangle ^\Box$ & \shortstack[l]{$\SI{8403(5)}{\hertz}$ \\ $\SI{-467(26)}{\hertz}$}
        & \shortstack[m]{$V_m(t) = \mathsf{C}(t, \tau) \cdot \varepsilon_q V_0\cos\left(\omega_{q} t + \frac{\omega_{c} \tau}{\pi^2}\left( \sin\left(\frac{\pi t}{2\tau}\right)-1\right)\right)\cdot \mathbf{1}_{\left[0, \tau \right]}$ \\
        $+~\varepsilon_q V_0\cos\left(\omega_{q} t\right)\cdot \mathbf{1}_{\left[\tau, \tau' \right]}$ \\
        $x_m(t)=\varepsilon_lx_{0} \cos\left(\frac{\omega_{q}}{2}t\right) \cdot \mathbf{1}_{\left[\tau, \tau'\right]}$}
        & \shortstack[l]{$\omega_{q}/2\pi=\SI{16.4}{\kilo\hertz}$ \\ $\varepsilon_q=50~\%$ \\ $\tau=\SI{5}{\milli\second}$ \\ $\omega_{c}/2\pi=\SI{942.478}{\hertz}$ \\ $\varepsilon_l=5~\%$ \\
        $\tau' = \SI{853.7}{\micro\second}$ } \\
    \midrule
    $\left|\mathcal{C}_{1.1}^{+} \right \rangle ^\diamond$ & \shortstack[l]{$\SI{8286(5)}{\hertz}$ \\ $\SI{-467(26)}{\hertz}$}
        & \shortstack[m]{$V_m(t) = \mathsf{C}(t, \tau) \cdot \varepsilon_q V_0\cos\left(\omega_{q} t + \frac{\omega_{c} \tau}{\pi^2}\left( \sin\left(\frac{\pi t}{2\tau}\right)-1\right)\right)\cdot \mathbf{1}_{\left[0, \tau \right]}$ \\
        $+~\varepsilon_q V_0\cos\left(\omega_{q} t\right)\cdot \mathbf{1}_{\left[\tau, \tau' \right]}$ \\
        $x_m(t)=\varepsilon_lx_{0} \cos\left(\frac{\omega_{q}}{2}t\right) \cdot \mathbf{1}_{\left[\tau, \tau'\right]}$}
        & \shortstack[l]{$\omega_{q}/2\pi=\SI{16.4}{\kilo\hertz}$ \\ $\varepsilon_q=50~\%$ \\ $\tau=\SI{5}{\milli\second}$ \\ $\omega_{c}/2\pi=\SI{942.478}{\hertz}$ \\ $\varepsilon_l=5~\%$ \\
        $\tau' = \SI{853.7}{\micro\second}$ } \\
    \midrule
    $\left|\mathcal{C}_{1.0}^{+} \right \rangle ^\triangle$ & \shortstack[l]{$\SI{8166(5)}{\hertz}$ \\ $\SI{-467(26)}{\hertz}$}
        & \shortstack[m]{$V_m(t) = \mathsf{C}(t, \tau) \cdot \varepsilon_q V_0\cos\left(\omega_{q} t + \frac{\omega_{c} \tau}{\pi^2}\left( \sin\left(\frac{\pi t}{2\tau}\right)-1\right)\right)\cdot \mathbf{1}_{\left[0, \tau \right]}$ \\
        $+~\varepsilon_q V_0\cos\left(\omega_{q} t\right)\cdot \mathbf{1}_{\left[\tau, \tau' \right]}$ \\
        $x_m(t)=\varepsilon_lx_{0} \cos\left(\frac{\omega_{q}}{2}t\right) \cdot \mathbf{1}_{\left[\tau, \tau'\right]}$}
        & \shortstack[l]{$\omega_{q}/2\pi=\SI{16.4}{\kilo\hertz}$ \\ $\varepsilon_q=50~\%$ \\ $\tau=\SI{5}{\milli\second}$ \\ $\omega_{c}/2\pi=\SI{942.478}{\hertz}$ \\ $\varepsilon_l=5~\%$ \\
        $\tau' = \SI{853.7}{\micro\second}$ } \\
    \midrule
    $\left|1 \right \rangle ^\Box$ & \shortstack[l]{$\SI{8403(5)}{\hertz}$ \\ $\SI{-467(26)}{\hertz}$}
        & \shortstack[m]{$x_m(t)=\varepsilon_lx_{0} \cos\left(\omega_{l}t\right) \cdot \mathbf{1}_{\left[0, \tau\right]}$}
        & \shortstack[l]{$\omega_{l}/2\pi=\SI{8.2}{\kilo\hertz}$ \\ $\varepsilon_l=2.5~\%$ \\ $\tau=\SI{5}{\milli\second}$ }\\ 
    \midrule
    $\left|1 \right \rangle ^\diamond$ & \shortstack[l]{$\SI{8286(5)}{\hertz}$ \\ $\SI{-467(26)}{\hertz}$}
        & \shortstack[m]{$x_m(t)=\varepsilon_lx_{0} \cos\left(\omega_{l}t\right) \cdot \mathbf{1}_{\left[0, \tau\right]}$}
        & \shortstack[l]{$\omega_{l}/2\pi=\SI{8.2}{\kilo\hertz}$ \\ $\varepsilon_l=2.5~\%$ \\ $\tau=\SI{5}{\milli\second}$ }\\ 
    \end{tabular}
\end{table*}

\begin{table*}[htbp]
    \setlength{\tabcolsep}{3pt} 
    \centering
    \small
    {\fontsize{8.5}{9.5}\selectfont
    \noindent Trap and modulation parameters (continued)\par
    }
    \vspace{5pt}
    \begin{tabular}{
        >{\centering\arraybackslash}p{0.12\textwidth}
        >{\centering\arraybackslash}p{0.15\textwidth}
        >{\centering\arraybackslash}p{0.54\textwidth}
        >{\raggedright\arraybackslash}p{0.15\textwidth}
    }
    \toprule
    \textbf{State} & \textbf{\shortstack{Trap parameters \\ $\omega/2\pi$ \\ $2K/2\pi$} }& \textbf{Modulation function} & \textbf{Modulation parameters}\\
    \midrule
    $\left|1 \right \rangle ^\triangle$ & \shortstack[l]{$\SI{8166(5)}{\hertz}$ \\ $\SI{-467(26)}{\hertz}$}
        & \shortstack[m]{$x_m(t)=\varepsilon_lx_{0} \cos\left(\omega_{l}t\right) \cdot \mathbf{1}_{\left[0, \tau\right]}$}
        & \shortstack[l]{$\omega_{l}/2\pi=\SI{8.2}{\kilo\hertz}$ \\ $\varepsilon_l=2.5~\%$ \\ $\tau=\SI{5}{\milli\second}$ }\\ 
    \midrule
    $\left|1 \right \rangle ^\heartsuit$ & \shortstack[l]{$\SI{8286(5)}{\hertz}$ \\ $\SI{-467(26)}{\hertz}$}
        & \shortstack[m]{$V_m(t) = \mathsf{C}(t, \tau) \cdot \varepsilon_q V_0\cos\left(\omega_{q} t + \frac{\omega_{c} \tau}{\pi^2}\left( \sin\left(\frac{\pi t}{2\tau}\right)-1\right)\right)\cdot \mathbf{1}_{\left[0, \tau \right]}$ \\
        $+~\varepsilon_q V_0 \cos\left(\omega_{q} t\right)\cdot \mathbf{1}_{\left[\tau, \tau' \right]}$ \\ 
        $+~\mathsf{C}(t, \tau)\cdot\varepsilon_q V_0\cos\left(\omega_{q} t + \frac{\omega_{c} \tau}{\pi^2} \sin\left(\frac{\pi}{2} \left(1+\frac{t}{\tau} \right) \right)\right)\cdot \mathbf{1}_{\left[\tau', \tau'+\tau \right]}$ \\
        $x_m(t)=\varepsilon_lx_{0} \cos\left(\frac{\omega_{q}}{2}t\right) \cdot \mathbf{1}_{\left[\tau, \tau'\right]}$}
        & \shortstack[l]{$\omega_{q}/2\pi=\SI{16.4}{\kilo\hertz}$ \\ $\varepsilon_q=50~\%$ \\ $\tau=\SI{5}{\milli\second}$ \\ $\omega_{c}/2\pi=\SI{942.478}{\hertz}$ \\ $\varepsilon_l=5~\%$ \\
        $\tau' = \SI{853.7}{\micro\second}$ } \\
    \midrule
    
    $\left|1 \right \rangle$
        & \shortstack[l]{$\SI{7400(1)}{\hertz}$ \\ $\SI{-356(3)}{\hertz}$}
        & $x_m(t) = \mathcal{G}(t, \tau, \sigma) \cdot \varepsilon_lx_{0} \sin\left(\omega_{l}t\right)  \cdot \mathbf{1}_{\left[0, \tau \right]}$
        & \shortstack[l]{$\varepsilon_{l} = 4.5~\%$ \\ $\omega_{l}/2\pi=\SI{7.6}{\kilo\hertz}$ \\ $\sigma=\SI{2.5}{\milli\second}$ \\ $\tau=\SI{5}{\milli\second}$} \\
    \midrule
    $\propto \left|0 \right \rangle + \left|1 \right \rangle$
        & \shortstack[l]{$\SI{7400(1)}{\hertz}$ \\ $\SI{-356(3)}{\hertz}$}
        & $x_m(t) = \mathcal{G}(t, \tau, \sigma) \cdot \varepsilon_lx_{0} \sin\left(\omega_{01}t\right) \cdot \mathbf{1}_{\left[0, \tau \right]}$
        & \shortstack[l]{$\varepsilon_{l} = 2.3~\%$ \\ $\omega_{l}/2\pi=\SI{7.55}{\kilo\hertz}$ \\ $\sigma=\SI{3}{\milli\second}$ \\ $\tau=\SI{6}{\milli\second}$} \\
    \midrule
    $\ket{2}$
        & \shortstack[l]{$\SI{7400(1)}{\hertz}$ \\ $\SI{-356(3)}{\hertz}$}
        & $V_m(t) = \varepsilon_q V_0 \sin(\omega_{q} t) \cdot \mathbf{1}_{\left[0, \tau \right]}$
        & \shortstack[l]{$\omega_{q}/2\pi=\SI{15.2}{\kilo\hertz}$ \\ $\varepsilon_q=9~\%$ \\ $\tau=\triangle{\text{2}}\,\SI{}{\milli\second}$} \\
    \midrule
    $\propto\ket{0}+\ket{2}$
        & \shortstack[l]{$\SI{7400(1)}{\hertz}$ \\ $\SI{-356(3)}{\hertz}$}
        & $V_m(t) = \varepsilon_q V_0\sin(\omega_{q} t) \cdot \mathbf{1}_{\left[0, \tau \right]}$
        & \shortstack[l]{$\omega_{q}/2\pi=\SI{15.2}{\kilo\hertz}$ \\ $\varepsilon_q=5.5~\%$ \\ $\tau=\SI{2}{\milli\second}$} \\
    \bottomrule
    \end{tabular}
    
    \par\small
    \vspace{0.2cm}
    Envelope functions:\\
    Square pulse: $\mathbf{1}_{\left[a, b \right]}(t) = H\left(t-a \right) - H\left(t-b \right)$ where $H$ is the Heaviside step function\\
    Gaussian: $\mathcal{G}(t, \tau, \sigma) = e^{-\frac{(t-\tau/2)^2}{2\sigma^2}} $ \\
    raised cosine: $\mathsf{C}(t, \tau) = \tfrac{1}{2}\!\left[1 - \cos\!\left(\tfrac{\pi t}{\tau}\right)\right]$ \\
    \footnotesize
    \vspace{0.15cm}
    \begin{tabular}{@{}l@{}}
    $^{\Box}$ Resonantly prepared state.\\
    $^{\Diamond}$ Prepared at a depth corresponding to a 100 Hz detuning.\\
    $^{\triangle}$ Prepared at a depth corresponding to a 200 Hz detuning.\\
    $^{\heartsuit}$ $\ket{0} \rightarrow \ket{C^+_\alpha} \rightarrow \ket{C^-_\alpha} \rightarrow \ket{1}$ protocol.
    \end{tabular}
\end{table*}

\end{document}